\documentclass[12pt,preprint]{aastex}

\newcommand{\um}{${\rm \mu m~}$}
\newcommand{\umns}{${\rm \mu m}$}
\newcommand{\mh}{${\rm H_2~}$}
\newcommand{\mhns}{${\rm H_2}$}
\newcommand{\lily}{IRAS 16594$-$4656~}
\newcommand{\lilyns}{IRAS 16594$-$4656}

\newcommand{\henns}{Hen 3$-$401}
\newcommand{\rob}{Rob~22~}
\newcommand{\robns}{Rob~22}

\slugcomment{Accepted for publication in the Astrophysical Journal}
\shorttitle{H$_2$ IN 3 BIPOLAR PPNs}
\shortauthors{Hrivnak et al.}

\begin{document}


\title{A STUDY OF H$_2$ EMISSION IN THREE BIPOLAR PROTO-PLANETARY
NEBULAE: IRAS 16594$-$4656, HEN 3$-$401, AND ROB
22\altaffilmark{1,2} }


\author{Bruce J. Hrivnak\altaffilmark{3}, Nathan Smith\altaffilmark{4}, Kate Y. L. Su\altaffilmark{5}, and Raghvendra Sahai\altaffilmark{6} }

\altaffiltext{1}{This work was based on observations with the
NASA/ESA Hubble Space Telescope, obtained at the Space Telescope
Institute, which is operated by AURA, Inc., under NASA contract
NAS5-26555.}  \altaffiltext{2}{The paper is based on observations
obtained at the Gemini Observatory, which is operated by the
Association of Universities for Research in Astronomy, Inc., under
a cooperative agreement with the NSF on behalf of the Gemini
partnership: the National Science Foundation (United States), the
Particle Physics and Astronomy Research Council (United Kingdom),
the National Research Council (Canada), CONICYT (Chile), the
Australian Research Council (Australia), CNPq (Brazil) and CONICET
(Argentina).}

\altaffiltext{3}{Department of Physics and Astronomy, Valparaiso University,
Valparaiso, IN 46383, USA; bruce.hrivnak@valpo.edu}
\altaffiltext{4}{Department of Astronomy, University of California,
Berkeley, CA 94720-3411, USA; nathans@astro.berkeley.edu}
\altaffiltext{5}{Steward Observatory, The University of Arizona,
Tucson, AZ 85721, USA; ksu@as.arizona.edu}
\altaffiltext{6}{Jet Propulsion Laboratory, California Institute of Technology, MS 183-900, 
Pasadena, CA, 91109, USA; raghvendra.sahai@jpl.nasa.gov}

\begin{abstract}
We have carried out a spatial-kinematic study of three
proto-planetary nebulae, IRAS 16594$-$4656, Hen 3-401, and Rob 22.
High-resolution \mh images were obtained with NICMOS on the {\it HST} and
high-resolution spectra were obtained with the Phoenix spectrograph on Gemini-South.  
IRAS 16594$-$4656 shows a ``peanut-shaped'' bipolar structure with \mh
emission from the walls and from two pairs of more distant,
point-symmetric faint blobs.  The velocity structure shows the
polar axis to be in the plane of the sky, contrary to the
impression given by the more complex visual image and the
visibility of the central star, with an ellipsoidal velocity structure.  
Hen 3-401 shows the \mh emission
coming from the walls of the very elongated, open-ended lobes seen
in visible light, along with a possible small disk around the
star.  The bipolar lobes appear to be tilted 10$-$15$\arcdeg$ with
respect to the plane of the sky and their kinematics display a Hubble-like flow.  
In Rob 22, the \mh appears in the form of an ``S''-shape,
approximately tracing out the similar pattern seen in the visible.
\mh is especially seen at the ends of the lobes and at two
opposite regions close to the unseen central star.  The axis of
the lobes is nearly in the plane of the sky.  Expansion ages of
the lobes are calculated to be $\sim$1600 yr (IRAS 16594$-$4656),
$\sim$1100 yr (Hen 3-401), and $\sim$640 yr (Rob 22), based upon
approximate distances.
\end{abstract}




\section{INTRODUCTION}
\label{intro}

Proto-planetary nebulae (PPNs) are objects in transition between
the asymptotic giant branch (AGB) and planetary nebula (PN) stages
of stellar evolution.  Studies over the past decade, particularly
imaging studies with the {\it Hubble Space Telescope} ({\it HST}),
have shown that this transition stage is key to understanding the
shaping of PNs \citep{balick02}.  In the interacting stellar winds
model \citep{kwok82}, which was later generalized to include an
equatorial density gradient \citep{balick87}, it was assumed that
a fast wind from the central star interacted with the detached,
slowly-moving remnant AGB envelope to shape the nebula.  {\it HST} images of
PPNs show that many and perhaps most of them display a basic
bipolar structure \citep{ueta00,su01,sahai07,siod08}. These are shapes
that appear to be further developed in the PN stage. In addition,
a point symmetry is often seen; \citet{sahai98a} attribute this to
collimated jets, which may be episodic and change their direction,
and which they advocate as the main shaping agents of PNs.

Studying \mh in PPNs is one of the best ways to probe the presence
of a fast, perhaps collimated, post-AGB wind and its effect in shaping the nebula.  
\mh is the main constituent of the detached circumstellar envelope around a
PPN. 
%and it can be shock-excited by collisions from the wind.
While a fast wind (V$>$100 km s$^{-1}$) has the energy to dissociate 
the \mh on direct impact, it can also produce shocks in the medium that move 
at a slower speed and collisionally excite the \mh.
Recent surveys have shown the presence of shock-excited \mh in
bipolar PPNs \citep{garher02, kelly05}. Detailed high-resolution
\mh studies have been published of approximately a half dozen PPNs
\citep{sahai98, kastner01, cox03, davis05, hri06, vds08}.

In this paper we present the results of a detailed \mh emission
study of three bipolar PPNs, IRAS 16594$-$4656 (``Water Lily
Nebula''), Hen 3$-$401 (IRAS 10178$-$5958), and Rob 22 (IRAS
10197$-$5750).  All three have relatively hot central stars for
PPNs, B and A spectral types, and all possess large infrared excesses
due to circumstellar dust \citep{partha89,garlar99a}.  
All three are or would be classified as DUPLEX (``DUst Prominent,
Longitudinally-EXtended'') nebulae in the classification system of 
\citet[see also \citet{siod08}]{ueta00}.

Rob 22 has a nebula that looks like a pair of butterfly wings.  At
high spatial resolution, the lobes show a filamentary structure in
visible light.  In addition to bipolar structure, some point
symmetry is apparent in the lobes.  A large halo extends out to
25$\arcsec$ \citep{sahai99a}. The nebula appears to be viewed
essentially edge on, with a dark lane obscuring the star in
visible light. 
In the more detailed morphological classification system of \citet{sahai07},
this object is classified as Bcw,ml,ps(s),h(e): bipolar nebula with closed lobes, 
central obscuring waist, minor lobes present, point-symmetric shape, and 
enlongated halo.
The spectral type is A2~Ie, based on light reflected
off the lobes of the nebula \citep{allen78}. The circumstellar
envelope appears to have a dual chemistry, with an oxygen-rich
region evidenced by OH maser emission 
\citep[OH284.18-0.79;][]{allen80} and crystalline silicate emission \citep{mol02}
and a carbon-rich region evidenced by PAH emission \citep{mol97}.

The nebula of Hen 3-401 possesses long, narrow lobes.  It seems to
be viewed at a slightly larger angle with respect to the plane of
the sky, with the central star appearing faintly between the lobes
\citep{sahai99b}. 
The morphological classification of this object is Bow*(0.6),sk: 
bipolar nebula with open lobes, central obscuring waist with star evident 
(at 0.6 $\mu$m), with a skirt-like structure \citep{sahai07}.
The spectral type of the star is Be, and the
spectrum shows strong Balmer emission lines and lots of permitted
and forbidden low-excitation emission lines \citep{allen78,
garlar99b}. The circumstellar envelope appears to be carbon-rich,
since it shows CO emission \citep{loup90, buj91} and PAH emission
\citep{partha01} but not OH \citep{silva93} emission or
crystalline silicates \citep{partha01}.

In IRAS 16594$-$4656, the star is clearly seen and relatively
bright compared to the nebula, giving the impression that the
nebula is at a larger orientation angle. 
The nebula has the appearance of a basic bipolar structure with a pronounced 
point symmetry, consisting of three pairs of oppositely-directed, slightly 
curved, thin lobes.  Also seen is a smaller elliptical structure oriented from northwest to 
southeast \citep{hri99}.
The morphological classification of this object is 
Mcw*(0.6),an,ps(m,an),h(a): multipolar nebula with closed lobes, 
central obscuring waist with star evident (at 0.6 $\mu$m), ansae present, 
point symmetry with two or more pairs of diametrically opposed lobes and 
with ansae, and a halo with centrosymmetric arcs \citep{sahai07}.
Its spectrum shows
Balmer emission lines \citep{garlar99a} and has been classified as
B7 based on the H$\alpha$ line \citep{vds00}.  \citet{rey02}, on
the basis of high-resolution spectra, fit the spectrum with
T$_{\rm eff}$=14,000 K and log{\it g}=2.1, slightly hotter than
$\beta$ Ori (B8~I); this indicates a spectral type of B5-7 Ie. He
also observed many emission lines.   The nebula appears to be
carbon rich, showing PAH features and displaying the 21 $\mu$m and
30 $\mu$m features seen in carbon-rich evolved stars
\citep{garlar99a,volk02}.
\citet{vds08} recently published a detailed kinematic and morphologic study 
of this object using H$_2$ emission, and we will refer to their results in the present 
study.

We begin with an examination of the high-resolution \mh images
(see \citet{sahai00} for preliminary images of Rob 22 and Hen
3$-$401 and \citet{hri04} for preliminary images of IRAS 16594$-$4656), then
present new high-resolution, long-slit \mh spectra, and then
follow with a discussion of the kinematics. From these, we are
able to determine properties of the collimated winds in these
three PPNs and how they have shaped the surrounding nebulae.

\section{HIGH-RESOLUTION H$_2$ IMAGING OBSERVATIONS}
\label{h2 obs}

High-resolution near-infrared imaging observations of the three
PPNs were obtained with the {\it Hubble Space Telescope} ({\it
HST}) and the Near-Infrared Camera and Multiobject Spectrometer
(NICMOS) camera 2 (NIC2).  The camera has a resolution of
$0\farcs076\times0\farcs075~{\rm pixel^{-1}}$ and a field of view
of 19\farcs5 $\times$ 19\farcs3. Observations were made with a
narrow-band ${\rm H_2}$ filter (F212N) and also with an adjacent
continuum filter (F215N) for the purpose of distinguishing the \mh
1$-$0 S(1) emission from the scattered starlight.

Hen 3-401 and Rob 22 were observed on 1998 under General Observer
program ID No. 7840 (PI: Kwok) using the original {\it HST} NICMOS
camera. Observations were also made with the F160W (H) and F222M
(K) filters for both of these objects.  IRAS 16594$-$4656 was
observed in 2002 under General Observer program ID No. 9366 (PI:
Hrivnak) using the refurbished {\it HST} NICMOS camera.
Observations were also made with the F110W (J) filter.   The
observing log is listed in Table \ref{obslog}.

\placetable{obslog}

The data were taken in the multiple non-destructive mode
(MULTIACCUM) to obtain a high dynamic range without saturating the
detector. 
Three sets of observations were made through each filter at each position 
on the array, and each target was imaged at three (Hen 3-401, Rob 22) or
four (IRAS 16594-4656) different positions of the NIC2 array using a 
predefined dither pattern.

The data were processed using the standard NICMOS {\it calnica} pipeline, which applied bias subtraction, dark subtraction, flat-fielding, and cosmic ray removal. 
The more recent data for IRAS 16594$-$4656 were re-calibrated using the HST archival On-the-Fly Reprocessing system, which provides recalibration using the most
up-to-date reference files. 
The individual files were then corrected for the ``pedestal effect'' 
using IRAF/STSDAS {\it pedsub} task 
to flattening the sky value in the four quadrants of the NIC2 array.  
We then combined the twelve images with
each filter using the IRAF STSDAS/Dither package with the parameters
suggested in the {\it HST Dither Handbook} \citep{koekemoer02}.  
The data for Hen3$-$401 and Rob 22, which were taken and reduced earlier, 
were reduced with the same philosophy but in a more manual fashion, as 
described by \citet{su03}.

\begin{figure}
\epsscale{0.7} \plotone{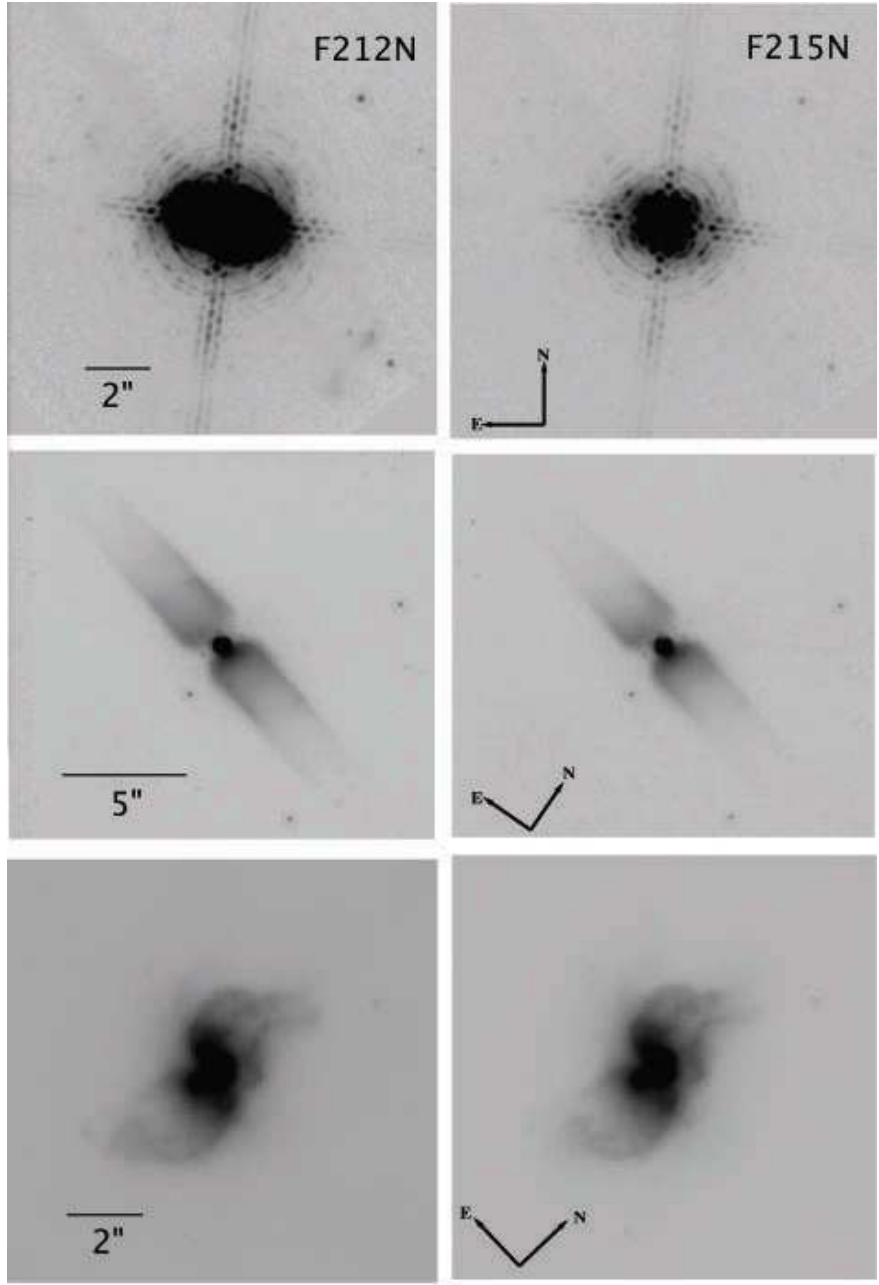}
\caption{The flux-calibrated F212N (left) and F215N (right) images
of IRAS 16594$-$4656 (top), Hen 3-401 (middle), and Rob 22
(bottom).  The images are all in reversed, logarithmic scale.  \label{h2_raw} }
\end{figure}

The reduced F212N and F215N images were then flux calibrated using
the appropriate calibration values for each filter before and
after the NICMOS refurbishing.  These calibration values are
listed in Table \ref{obscal}.  The individual F212N and F215N
images for the three PPNs, with the thermal backgrounds removed,
are displayed in Figure \ref{h2_raw}.

\placetable{obscal}

\placefigure{h2_raw}

For \lilyns, the relatively-bright central star is surrounded by a distinct 
peanut-shaped nebula in the F212N image, while only some smaller and 
fainter nebulosity surrounds the star in the F215N image. Thus
one can immediately see that most of the nebular light in the
F212N image is due to the \mh emission rather than scattered
starlight. 
One can also see 
%two faint clumps of emission southwest of the lobes in the F212N image. 
faint emission from two of the pairs of longer, thin lobes, with brighter emission 
(``ansae'') at their ends.
For \robns, the F212N and F215N images appear
rather similar. They each show, rather than a central star,
emission coming from two closely-spaced regions separated by a 
dark equatorial band. These bright
regions may represent light from the central star scattered toward
us from above and below an obscuring disk or torus.  The two
bright regions lie roughly along the line that connects the
approximately point-symmetric outer ends of the two lobes. The
dark space between the two bright regions might thus be due to the
torus. The nebula appears to be a little brighter in the F212N
image than in the F215N image, especially in the outer regions,
indicating that in the F212N image the nebula is seen mostly in
scattered light but with some contribution from \mh emission. For
\henns, the nebula is seen more clearly and is larger in the F212N
image than in the F215 image, indicating that in the F212N image,
it is seen in both scattered light and \mh emission. The star is
bright in each image.

\section{H$_2$ IMAGES}
\label{h2 images}

To produce images of these three PPNs showing only \mh emission
requires the removal of the scattered light from the F212N image,
which can be accomplished through the subtraction of an
appropriately scaled image of the adjacent continuum (F215N
image). The scaling is necessary to correct for differences in the
source continuum in the two bandpasses.  Differences due to the
filter profiles and detector response were corrected by the flux
calibration. In a similar previous study \citep{hri06}, we
examined the question of how best to scale the F215N continuum
image for subtraction from the F212N image to remove the scattered
continuum light.  The object of that study, IRAS 17150$-$3224, had
no other emission lines in these two bandpasses, and thus a simple
correction for the continuum levels was appropriate.  In these
cases, we also need to be concerned about other emission features
which may be contributing to the flux in the two filters.

In the following sub-sections we discuss the details of the
continuum removal and the resulting \mh image for each object.

\subsection{IRAS 16594$-$4656}

\begin{figure}
\epsscale{1.0}  \plotone{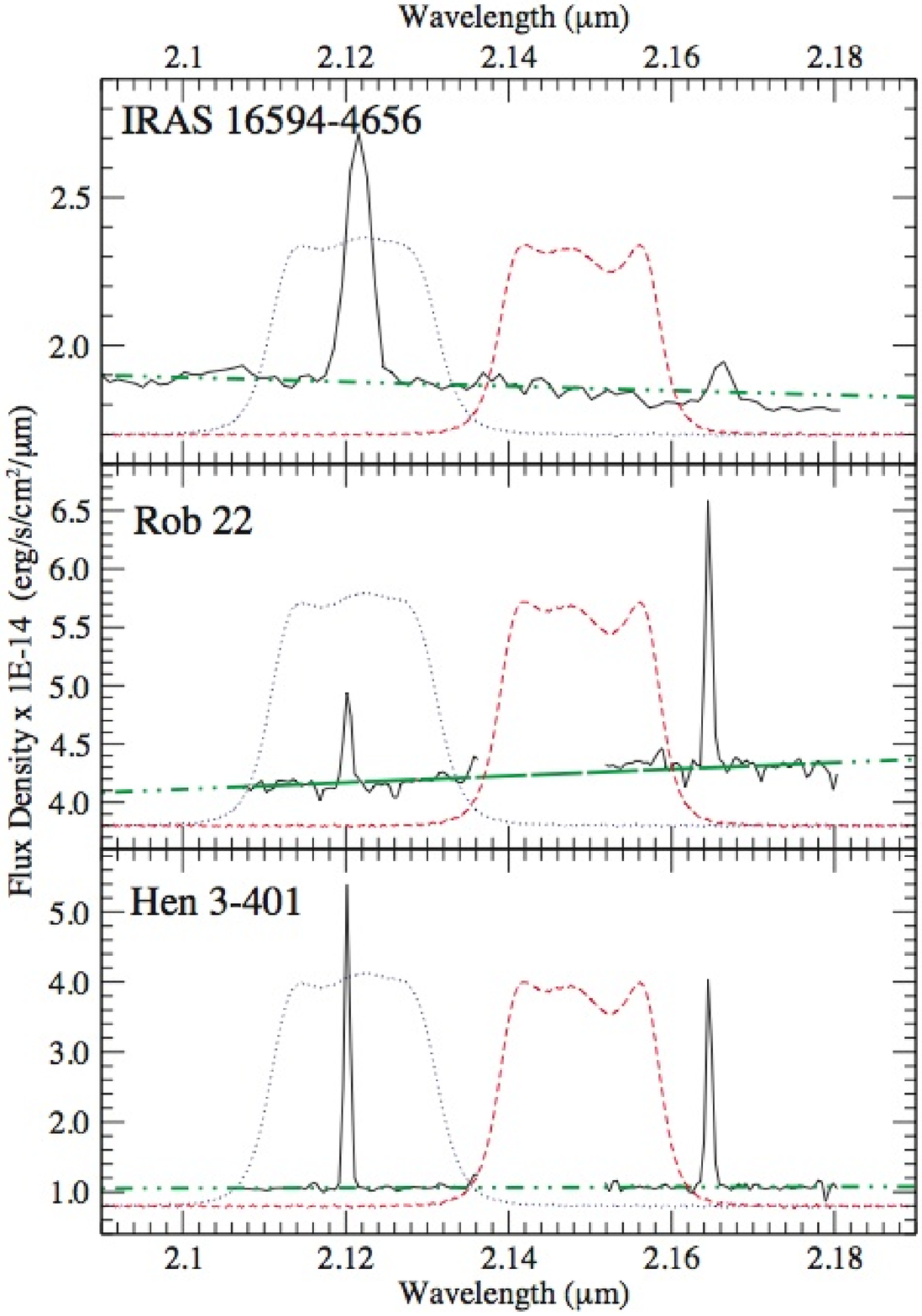}
%\plotone{fig2.eps}
\caption{The medium-resolution spectra of IRAS 16594$-$4656, Rob
22, and Hen 3-401 \citep{vds03,garher02}, showing the \mh and the
Br$\gamma$ emission features, along with the {\it HST} NICMOS
F212N and F215N filter profiles. The dotted and dashed lines show
the profiles of the F212N and F215N filters, respectively, plotted
in each case on an arbitrary scale.  The dashed-dotted line in
each panel shows the linear fit to the continuum. \label{filters} }
\end{figure}

A spectrum for IRAS 16594$-$4656 in the 2 $\mu$m region
(2.10$-$2.27 $\mu$m) has been published by \citet{vds03}.  It was
obtained with a long slit 1$\farcs$0 wide and oriented E-W.  The
spectrum shows a relatively strong \mh feature at 2.12 $\mu$m with
a much weaker Br$\gamma$ emission line.
%[H2/Br$\gamma$$\sim$7]
Since the Br$\gamma$ is at a wavelength that places it at the edge
of a wing of the F215N filter profile, its contribution is
insignificant in this image.  This is illustrated in Figure
\ref{filters}, which shows the spectrum of the object plotted
together with the filter profiles.   Thus, for the continuum
removal, we convolved a linear fit to the observed spectral
continuum (with the \mh emission removed) with the filter profiles
and the detector response using the STSDAS/SYNPHOT {\it calcphot}
task.  This resulted in a scale factor of 1.016 that we used to
scale the F215N image prior to subtraction of the flux-calibrated
images.

\placefigure{filters}

\begin{figure}
\epsscale{0.55}  \plotone{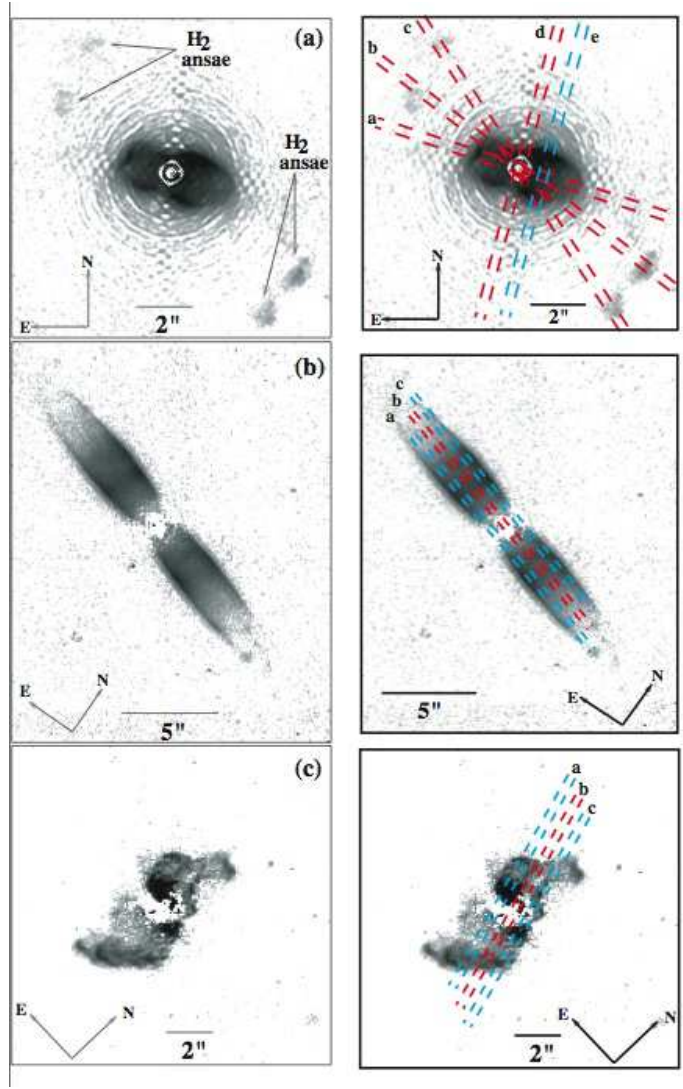}
%\plotone{Fig3.eps}
\caption{
In the left-hand panels are shown the resulting flux-calibrated \mh
images of IRAS 16594$-$4656 (top), Hen 3-401 (center), and Rob 22
(bottom).  In the right-hand panel of each is shown the \mh image
with the position of the Phoenix slits overlaid.  
The labelling of the slits corresponds to the
labelling in the position-velocity diagrams (Figs.
\ref{spec_16594}, \ref{spec_hen3}, \ref{spec_rob22}). \label{h2_images} }
\end{figure}

The resultant \mh image of IRAS 16594$-$4656 is displayed in
Figure \ref{h2_images}a.
It shows a limb-brightened, double-lobed bipolar structure
with the west lobe larger and brighter (ratio of 1.6) than the east lobe.
Approximate sizes (10$\sigma$ level) are 2$\farcs$6 by 2$\farcs$8 for the west lobe
and 2$\farcs$2 by 2$\farcs$5 for the east lobe.  In addition, one
can see two pairs of more distant \mhns-emitting 
%clumps on opposite sides of the star, 
ansae,
each pair collinear on a line passing
through the star. The %clumps 
ansae to the SW are brighter and farther
from the star than those to the NE, with distances of
6$\farcs$2 and 5$\farcs$4 for the %clumps 
ansae at PA=33$\arcdeg$ and
of 5$\farcs$8 and 4$\farcs$7 for the those at PA=54$\arcdeg$.
These %clumps 
ansae are found at the ends of %filaments %(``petals'')
the thin lobes  
seen in the visible light images \citep{hri99,su01}, and will be
discussed together with the visible image in Section \ref{discussion}. The
emission at the position of the star appears to approximately
cancel out, indicating that there is little, if any, \mh emission
from the star. 

\placefigure{h2_images}

\subsection{Hen 3-401}

Spectra of the 2.12 $\mu$m \mh line and the Br$\gamma$ line of Hen
3-401 have been published by \citet{garher02}.  They observed this
object twice with a 4$\farcs$5-wide slit oriented at P.A. =
70$\arcdeg$ and 90$\arcdeg$, approximately along the bipolar axis,
and they found the \mh emission to be clearly extended. The \mh
line is stronger than the Br$\gamma$ line by about a factor of
1.5, and since the latter lies in the wing of the F215N filter,
its contribution to the flux in the F215N image is very small.
This is illustrated in Figure \ref{filters}. Neglecting the
contribution of the line results in a scale factor of 0.994 (the
spectrum is essentially flat in this region).

The resulting \mh image of Hen 3-401 is shown in Figure  \ref{h2_images}b. 
\mh emission extends along the two almost cylindrical, slightly
barrel-shaped lobes, each with a size of 2$\farcs$5 $\times$
10$\arcsec$.  The east lobe is brighter, with a ratio of 1.2.  
The two lobes are limb-brightened and appear to be
brightest 2$\farcs$5$-$5$\farcs$3 (E) and 1$\farcs$5$-$4$\farcs$3
(W) from the star.  The ends of the lobes are open. Little, if
any, of the \mh emission appears to come from the central
star.\footnote{\citet{garlar99b} previously displayed the {\it
HST} \mh data, using a scale factor based upon the field stars,
and obtained a similarly appearing \mh image.}

\subsection{Rob 22}

Rob 22 was also observed spectroscopically in this region by
\citet{garher02}.  They observed with the slit oriented at P.A. =
90$\arcdeg$, in this case almost perpendicular to the bipolar
lobes. With a slit width of 4$\farcs$5, they likely included most
of the nebula but missed some of the \mh emission from the very
ends of the lobes.  Their spectra show the Br$\gamma$ line to be
stronger than the H$_2$ 2.12 $\mu$m \mh line by about a factor of 2.5,
but since it still falls on the extreme edge of the filter
profile, its contribution will be small. Nevertheless, we examined
the scaling of the F215N image with and without consideration of
this emission line. Neglecting the contribution of the line
results in a scale factor of 0.979, while including it results in
a scale factor of 0.967. We will use the scale factor with the
line included, but note that the difference is only at the 1\%
level.  Note, however, that the difference can be larger in localized 
regions, since the Br$\gamma$ and H$_2$ emissions likely have 
different distributions.

The resulting \mh image of Rob 22 is shown in Figure  \ref{h2_images}c.
The nebula appears to have somewhat of a point-symmetric ``S''
shape, with a bright clump in the inner part of each lobe near the
position of the unseen central star (clump to the NE is
brighter) and a ridge of \mh along the the end of each lobe
(especially the S lobe).
The north lobe is brighter, with a ratio of 1.2.
Much of the extended structure within a few arcsec of the star disappears after continuum subtraction, indicating that the features at the bases of the polar lobes nearest the star are due to scattered light from a dusty flared disk or torus, not H$_2$ emission.

\section{HIGH-RESOLUTION \mh SPECTRA}
\label{h2 spectra}

\subsection{Observations and Reduction}

High dispersion, long-slit spectral observations were made of the
three sources in the \mh $v$=1$-$0 S(1) line at 2.1218 $\mu$m
using the near-infrared echelle spectrograph Phoenix
\citep{hinkle98} on the 8-m Gemini-South telescope on Cerro
Pachon, Chile. Phoenix has a 1024$\times$256 InSb detector with a
pixel scale of 0$\farcs$085 $\times$ 1.26 km s$^{-1}$ at a
wavelength of $\sim$2~$\micron$. The slit width was 0$\farcs$34,
providing an effective velocity resolution of 6 km~s$^{-1}$
(R$\sim$50,000).  
The spectrum covers a range of $\sim$1300 km  s$^{-1}$ (0.0086 $\mu$m) 
centered on this \mh line.
The area of the sky projected by the long slit
onto the detector is $\sim$13$\farcs$4 in length, so for the
observations of IRAS~16594$-$4656 and Hen~3-401, more than one
exposure was taken moving along the slit, yielding a longer
effective slit to cover each target.  The observations were made
in 2003 under three queue observing programs (GS-2002B-Q-53,
GS-2003A-Q-27, GS-2003B-Q-41).
Image quality on the different nights varied from 0$\farcs$7 to
1$\farcs$0 as measured in the R band.

Several spectra were obtained for each of the PPNs at various
positions.  For each object, one spectrum was taken with the slit
oriented through the star along the major axis of the lobes.
Additional spectra were taken at various offsets and orientations.
For IRAS 16594$-$4656, two of the slit positions were oriented to
pass though the star and the pairs of \mh clumps.
An observing log listing the slit positions is given in Table
\ref{ph_obs}, and the orientations of the slit for these various
observations are shown in Figure \ref{ph_obs}.

\placetable{ph_obs}

The weather for these observations was mostly clear, although some
patchy clouds were present during observations of Rob 22 and IRAS
16594$-$4656.  Sky subtraction was accomplished by obtaining one
exposure of blank sky near each source immediately before or after
a pair of
exposures of the target.  Spectral flat fields were obtained using
an internal quartz emission lamp.  For each target, a bright
standard star was observed on the same night with the same grating
setting in order to correct for telluric absorption and for flux
calibration. Numerous telluric lines were also used for wavelength
calibration, adopting the telluric spectrum available from
NOAO\footnote{ftp://ftp.noao.edu/catalogs/arcturusatlas/ir/}.  We
calculated the apparent velocities adopting a rest wavelength of
21218.356 \AA\ for the H$_2$ $v$=1--0 S(1) line (Bragg, Brault, \&
Smith 1982), and these velocities were then corrected to a
heliocentric reference frame. Uncertainty in the resulting
velocities is roughly $\pm$1 km s$^{-1}$, dominated by scatter in
the dispersion solution for the numerous telluric lines across the
observed wavelength range.

\subsection{Position-Velocity Diagrams}

\subsubsection{IRAS 16594$-$4656}

\begin{figure}
\epsscale{1.0}  \plotone{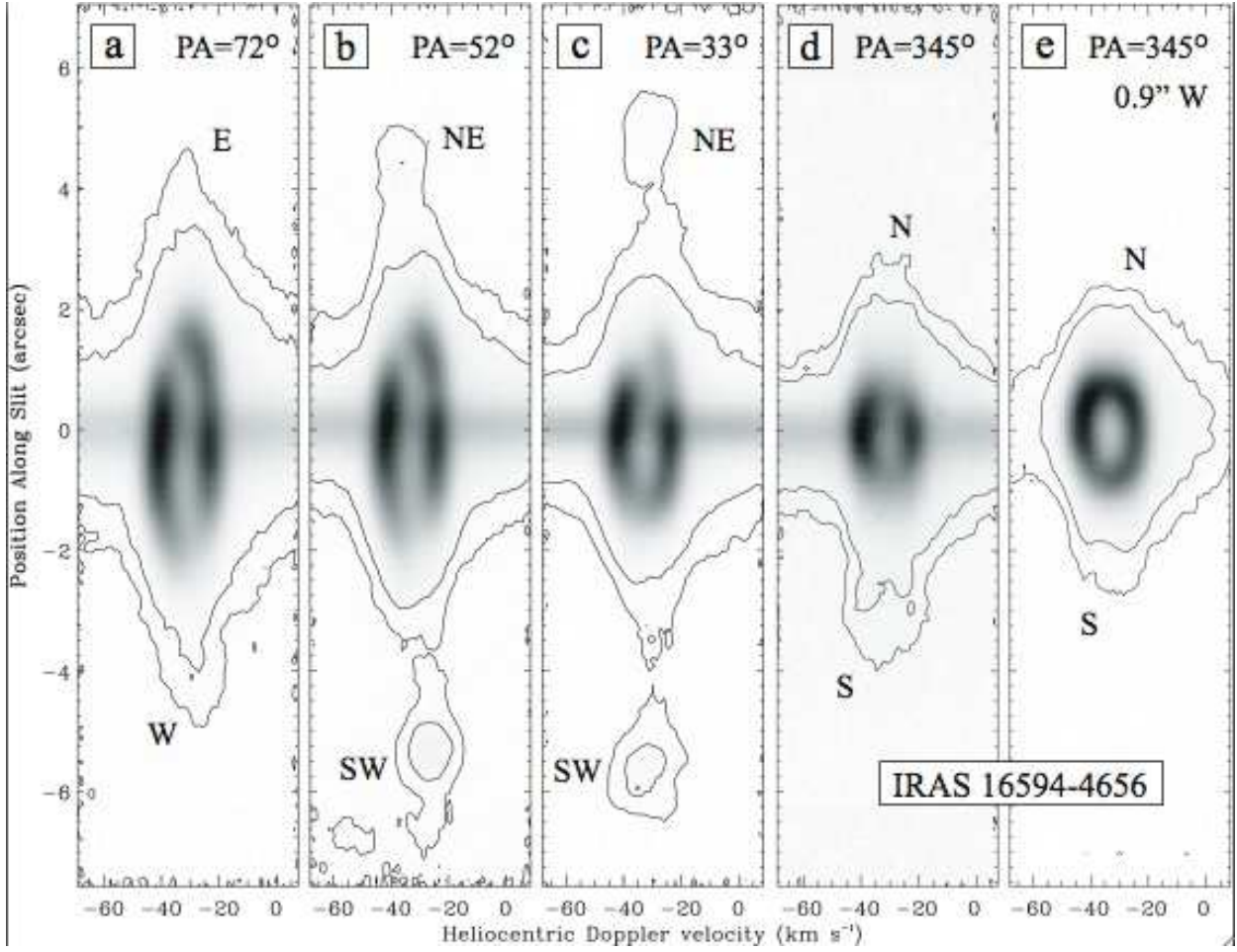}
%\plotone{fig.2Diras.eps}
\caption{The \mh position-velocity diagram of IRAS 16594$-$4656.
Panels a - e correspond to the five slit positions shown in Fig.
\ref{h2_images}. \label{spec_16594}}
\end{figure}

Spectra of IRAS 16594$-$4656 were obtained at five different
positions on the nebula, as illustrated in Figure 3.  The
resulting position-velocity (PV) diagrams are shown in Figure
\ref{spec_16594}. In all cases the brightest emission displays a
closed ellipse, but with different sizes and strong variations in
intensity around the ellipse.  In slit position {\it a} (PA=72$\arcdeg$), which is
along the major bipolar axis, and slit position {\it b}  (PA=52$\arcdeg$), which is
rotated 20$\arcdeg$ from the major axis, the inner ellipses seem
to be tilted slightly,
with the W-SW lobe slightly blue-shifted compared to the E-NE
lobe.  However, this is a small effect, and the polar axis appears to 
lie very close to the plane of the sky.

Slit positions {\it c}  (PA=33$\arcdeg$) and {\it d}  (PA=345$\arcdeg$) also pass through the star, but
are oriented closer to the minor axis (slit position d is along
the minor axis), and the size of the bright inner ellipse is
correspondingly smaller. Slit position {\it e}  (PA=345$\arcdeg$) is parallel to the minor
axis but offset from the star through the larger W lobe. The PV
diagram at this position is consistent with a slice through a
shell expanding with a velocity of $\sim$8-9 km s$^{-1}$ and with
the N side brighter.

\placefigure{spec_16594}

Intensity contours are shown in Figure \ref{spec_16594} to display
the faintest emission levels.  Slit position {\it a} shows that
the \mh emission extends $\pm$5\arcsec\ along the major axis. Slit
positions {\it b} and {\it c} pass through the clumps at the ends
of extended arms in the nebula, and in both spectra these clumps
are seen in \mh emission. In slit position {\it b}, the centroid
velocities of the two clumps differ by 10 km s$^{-1}$ ($-$36 to
$-$26), with each about 5 km s$^{-1}$ from the mean velocity of
$\sim$$-$30 km s$^{-1}$.  In slit position {\it c}, the velocity of
the two clumps is about the same. These velocities can be explained 
by a model in which the clumps are ejected by rotating jets. 
%At slit position {\it b}, the jet mechanism was tilted with
%respect to the plane of the sky and the brighter SW clump is
%moving away and the fainter NE clump is moving toward us, while at
%orientation {\it c}, the jets were pointed in the plane of the sky
%5and the clumps are moving perpendicular to our line of sight. 
The clumps at slit position {\it b} appear closer to the star than
those at slit position {\it c}; this can be a result of a
projection effect and/or temporal effect, assuming that they are
moving outward from the central star at the same speeds.  If the
clumps in each point-symmetric pair began moving from the location
of the star at the same time, then they should be at the same
distance from the star. However, in both cases the brighter SW
clumps appear farther from the star than the NE clumps, so the
assumption of constant velocity for the clumps may not be correct.
Alternatively, the clumps may simply be non-uniform brightness
enhancements at the ends of the filaments.

The bright ellipses are consistent with limb-brightened lobes.
When allowance is made for the resolution of the spectra, it appears 
that the interiors of the lobes are devoid of \mh emission.
In the four slit positions that include the star (Fig.
\ref{spec_16594}, panels $a-d$), the blue-shifted side of the
inner ellipse is brighter. This may be due to extinction within
the lobes or intrinsically brighter \mh emission from the side facing us.  
The elliptical morphologies of the PV diagrams for the
four slit positions through the star, especially the one along the
major axis, are consistent with an expanding ellipsoid.  Thus they
differ from the ``peanut-shape'' PV diagram that one might expect
from two bipolar expanding bubbles.  Thus the expansion along the
minor axis does not appear to be significantly retarded by a constraining
higher-density torus.

\begin{figure}
\epsscale{0.4}  \plotone{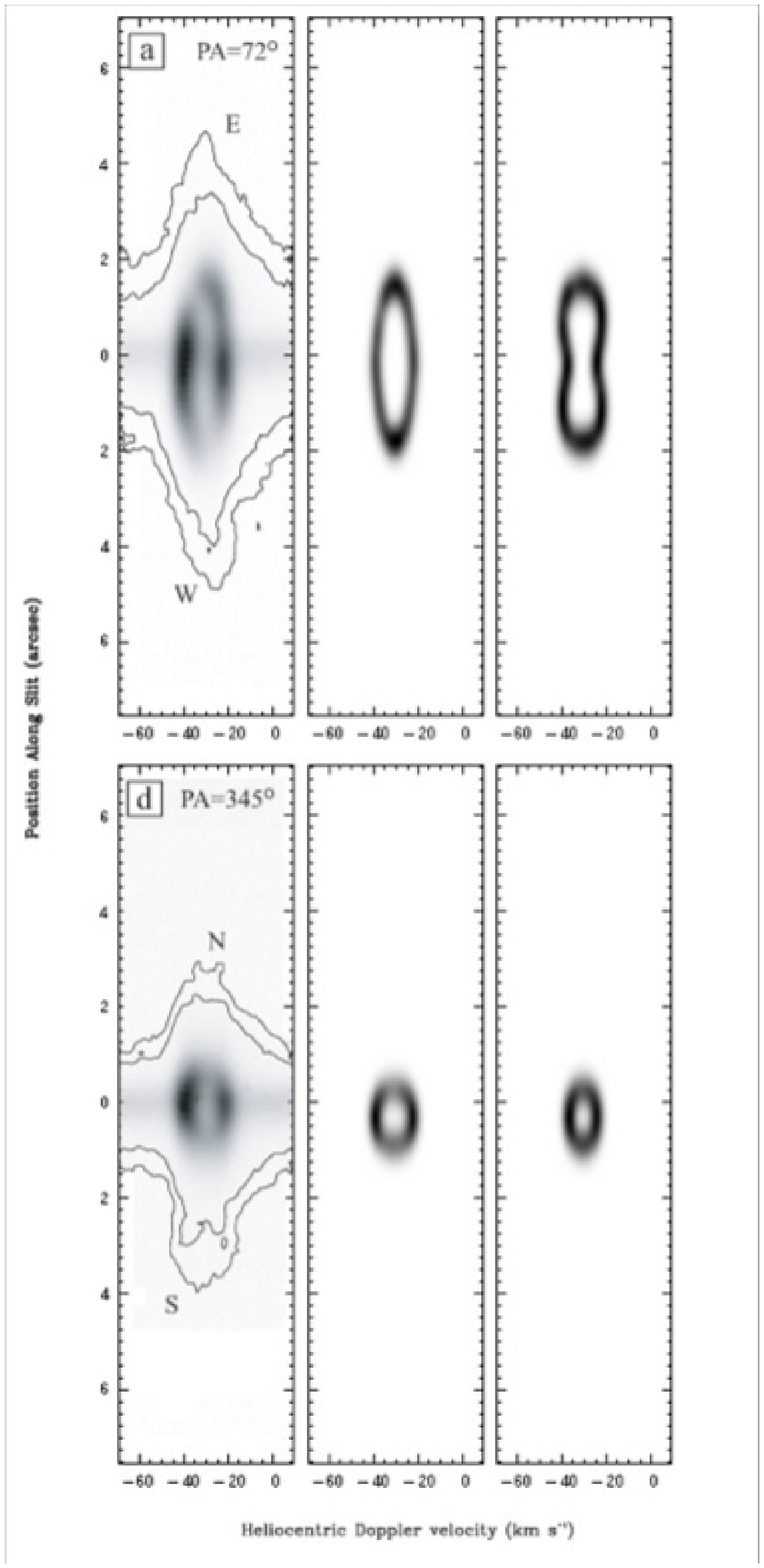}
%\plotone{SHAPE_figs.eps} 
\caption{PV diagram for IRAS 16594$-$4656. 
On the left side is shown the observed data, in the middle is shown the model results with a constant velocity plus slight Hubble flow ({\it V} = (8 + 2.4{\it r}/1$\arcsec$) km~s$^{-1}$), and on the right is shown the results with a pure Hubble flow ({\it V} = 12{\it r}/1$\arcsec$ km~s$^{-1}$).  The top row is for slit position {\it a} and the bottom row for slit postion {\it d} in 
Fig.~\ref{h2_images}.  Note the good agreement of the shape of the the middle panels (constant velocity with slight Hubble flow) with the observed data. (To fit the brightness would involve a modification in the surface emissivity from the constant value assumed.)  The model results have been convolved with the seeing and velocity resolution to produce the suitable images to compare with the observations. \label{SHAPE_16594} }
\end{figure}

We investigated the velocity field in more detail using the morpho-kinematic modeling
program {\it SHAPE} \citep{stef06}.
In this program, the morphology of the nebula and the velocity field can be specified 
and a resulting PV diagram produced.  The shape of the nebula was assumed to be
a three-dimensional symmetrical dumbbell based on the H$_2$ (cross-sectional) image.  
It was then assumed that the H$_2$ emissivity uniformly followed the contours of the nebula.
We began by investigating a velocity field in which the velocity increased linearly with distance 
(so-called ``Hubble flow''), which is seen in some PNs and is seen in Hen 3$-$401 (Section 4.2).  
This produced a PV diagram that had a peanut shape, contrary to the elliptical shape seen 
(see Figure \ref{SHAPE_16594}. left and right panels).  When we assumed on the other hand 
a constant velocity (9 km~s$^{-1}$), the PV diagram agreed more closely with the observed 
ellipse but was slightly diamond-shaped.  A very good fit was obtained by the addition of 
a slight Hubble flow, with {\it V} = (8 + 2.4{\it r}/1$\arcsec$) km~s$^{-1}$, as shown in 
the middle panels of Figure \ref{SHAPE_16594}.  
We determine, based on adding an inclination to the model, that the tilt of the polar 
axis is no more than $\sim$10$\arcdeg$ from the plane of the sky.

\placefigure{SHAPE_16594}

\begin{figure}
\epsscale{0.5}  \plotone{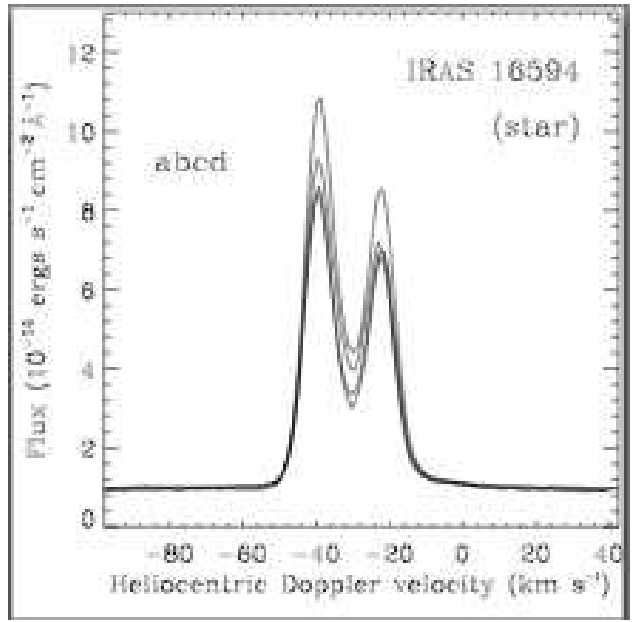}
%\plotone{starTraceIRAS.eps} 
\caption{Superimposed slices in
velocity at position along slit = 0\arcsec\ across the \mh
position-velocity diagram of IRAS 16594$-$4656.  (The differences
in flux levels likely arise from differences in seeing and
transparency.) \label{slice_16594} }
\end{figure}

Slices in velocity across the P-V diagrams at position along the
slit = 0\arcsec\ (along the line of sight through the star) with
width 0$\farcs$77 are displayed in Figure \ref{slice_16594}. These
show a uniform expansion of the shell with an expansion velocity
of V$_{\rm exp}$ = 9 km s$^{-1}$. The velocity center is at $-$30
km s$^{-1}$.  The expansion velocity of the lobe at 0$\farcs$9
west of the star is similar; this is consistent with uniform expansion of an
elliptical bubble.

\placefigure{slice_16594}

These observations reveal a systemic heliocentric radial velocity
of V$_{\rm hel}$(\mhns) = $-$30 km s$^{-1}$, which translates to
V$_{\rm LSR}$ = $-$26 km s$^{-1}$, with V$_{\rm exp}$ = 9 km
s$^{-1}$. These velocities can be compared with the
millimeter-wave CO $\it J$=1$-$0 emission line of V$_{\rm
LSR}$(CO) = $-$26 km s$^{-1}$ with V$_{\rm exp}$ = 16 km s$^{-1}$
\citep{loup90}. The CO expansion represents the velocity of the
outer remnant of the AGB mass loss expanding into the ambient ISM,
while the \mh expansion represents the velocity of the shock front
at the edge of the inner bipolar cavity along the line-of-site
direction, which is essentially perpendicular to the polar axis.
The difference between these two expansion velocities 
appears to reflect the two different kinematic regions.
Heliocentric radial velocity
measurements of individual lines in the visible spectrum showed a
range of $-$10 to $-$40 km s$^{-1}$, with average values of
$\sim$$-$20 km s$^{-1}$ for the absorption lines and $\sim$$-$32
km s$^{-1}$ for the emission lines \citep{rey02}.

\subsubsection{Hen 3-401}
\label{hen3 PV}

\begin{figure}
\epsscale{1.0}  \plotone{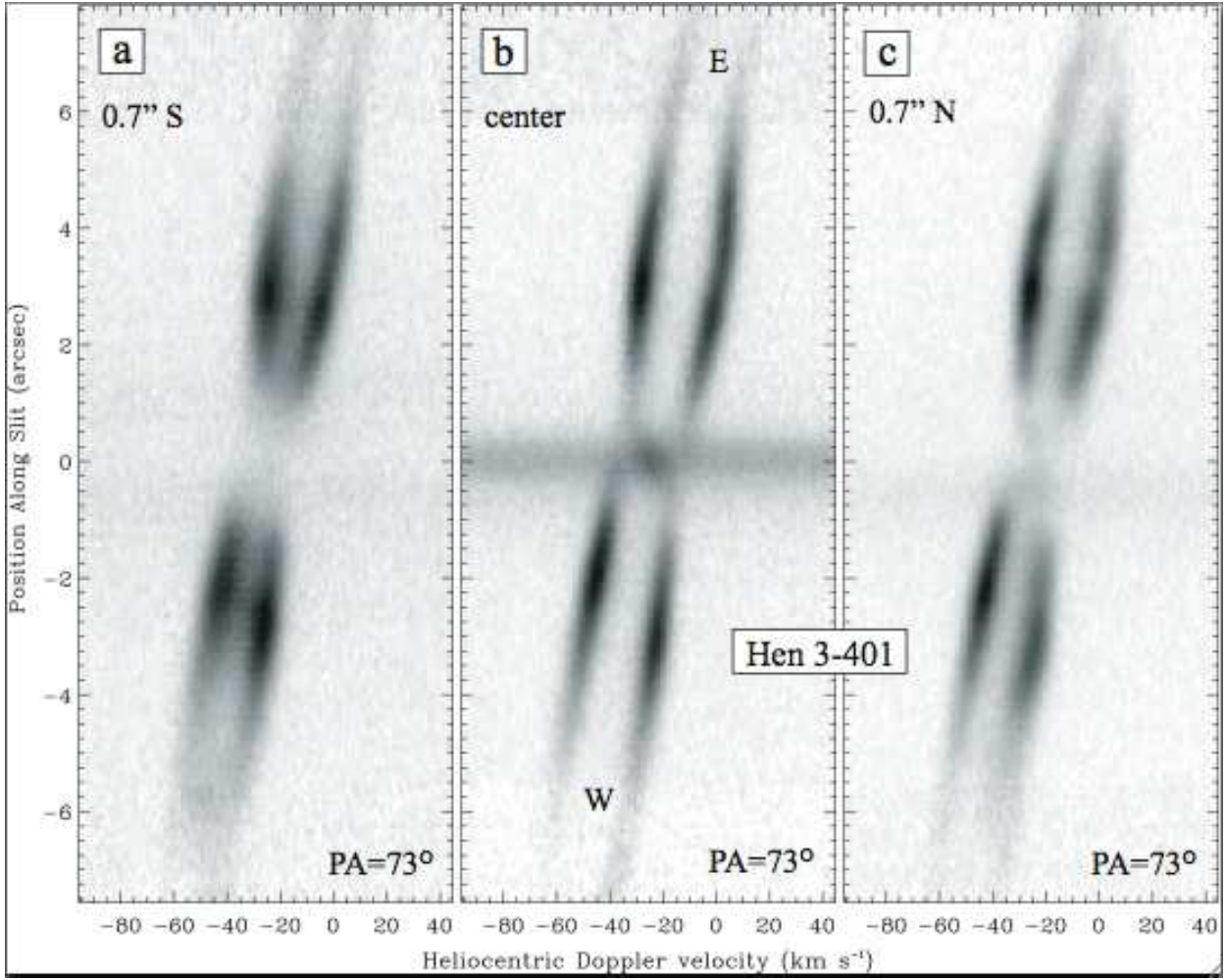}
%\plotone{fig.2Dhen.eps} 
\caption{The \mh position-velocity diagram
of Hen 3-401.  Panels a - c correspond to the three slit positions
shown in Fig. \ref{h2_images}.   \label{spec_hen3}}
\end{figure}

Spectra of Hen 3-401 were obtained at four different positions on
the nebula.  The resulting position-velocity diagrams for the
three oriented along the bipolar axis are shown in Figure
\ref{spec_hen3}. (The fourth one, passing through the star but oriented 
perpendicular to the axis, showed no
extended structure, only \mh emission at the position of the star
with the systemic velocity).  In each of these three slit
positions, we see a pair of convex arcs from each lobe. They are
not closed at the ends, indicating open lobes out to at least
$\pm$7\arcsec~ from the star. At the two offset positions, some
emission appears to partially fill the interior of the lobes; this
may simply be contamination from the very edges of the lobes due
to seeing.

\placefigure{spec_hen3}

The bipolar axis is obviously tilted from the plane of the sky,
with the W lobe moving toward us, and with velocities along the
lobes that are proportional to distance from the star (Hubble flow). 
The pairs of convex arcs represent emission from
the thin front and back walls of slightly barrel-shaped lobes. The
difference between the pair in each lobe is about 30 km s$^{-1}$
for the slit passing through the star (Fig. \ref{spec_hen3}b) but about 23
km s$^{-1}$ in the two offset positions; this is consistent with
seeing the radial expansion along the bipolar axis but only a
projected component of it in the offset positions.  If the
expansion really is a Hubble flow, then the relationship between
observed expansion velocities of the front and back sides of the
lobes in Figure \ref{spec_hen3}b compared to their apparent spatial size can
constrain the dynamical age of the nebula for a given distance.
If we assume axial symmetry --- so that the lobes have roughly the same 
length-to-width ratio as in H$_2$ images (this seems justifiable based in 
the similar morphology in Figures \ref{spec_hen3}b and \ref{hen3_multi}) 
--- then one can stretch the 
PV plot in Figure \ref{spec_hen3}b to have the correct aspect ratio.  
Figure \ref{spec_hen3}b thus represents a slice through the structure of the nebula in 
the plane along our line-of-sight \citep[see][for a more detailed 
explanation in the case of the bipolar lobes of Eta Carinae]{smith06}.  This yields 
a tilt of the polar axis from the plane of the sky for Hen 3-401 of 
roughly 15$\arcdeg$($\pm$5$\arcdeg$).  In that case, the lateral expansion 
speed of the lobes (the radial expansion speed perpendicular to the polar 
axis) is V$_{\exp}\approx$15.5($\pm$1) km s$^{-1}$.  
In the images, the polar lobes have a
width of roughly 2$\farcs$5, which corresponds to
3.7$\times$10$^{11}$($D_{kpc}$) km, where $D_{kpc}$ is the
heliocentric distance in kpc to Hen 3-401. Thus, the age of the
nebula would be only about 380 $\times D_{kpc}$ yr if the
expansion has been ballistic.  Assuming a distance of 3 kpc
\citep{buj91} thus yields a nebular age of 1100 yr.

\begin{figure}
\epsscale{0.5}  \plotone{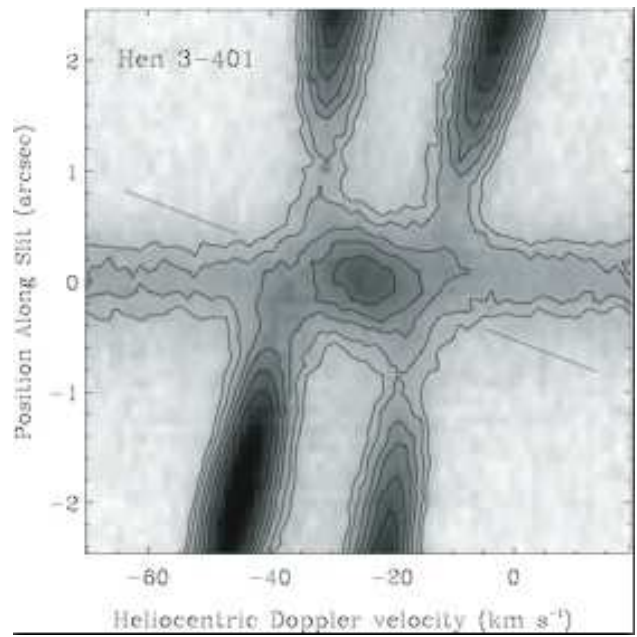}
%\plotone{fig.2DhenZOOM.eps} 
\caption{Enlarged \mh position-velocity diagram of the region around the central star of
Hen 3-401. \label{spec_hen3_zoom} }
\end{figure}

There is some weak \mh emission coming from the star as well.  In
fact, the very weak central peak appears to show some extended
structure, with a possible tilt $-$ i.e. it is slightly
blueshifted toward the east and redshifted toward the west, as
shown in Figure \ref{spec_hen3_zoom} in an expanded view of the
region around the central star. This may be emission from an inner
ring or disk surrounding the central star. This same type of
tilted structure from the central \mh peak is also seen in similar
spectra of the young PN M~2-9. In fact, it is remarkable that all
aspects of the overall kinematic structure of \mh emission from
Hen~3-401 in panel {\it b} of Figure~\ref{spec_hen3} are almost
identical to those of M 2-9 \citep{smith05b}. Even the lateral
expansion velocity of the lobes is roughly the same as for M 2-9.
Although the apparent size of Hen~3-401 is smaller by a factor of 
$\sim$4.5, it distance is larger by about the same amount, so the two objects
have the same size, nebular age, and overall appearance.  A 
detailed comparison of their central stars 
%, along with that of Mz~3 (Smith 2003), 
might therefore be quite illuminating.

\placefigure{spec_hen3_zoom}

At most positions, \mh emission from the far sides of the lobes is
fainter than on the near side, especially where the far side is
behind an area of bright emission on the near side.  A similar
effect is seen in M2-9, where it is probably due in part to
extinction by dust in the lobes \citep{smith05a}.  However, an
asymmetry does appear in the brightness of the east-west lobes of
Hen 3-401, with the brighter peak from the near side closer to the
star in the west lobe, position $\approx$ 2\arcsec\, while in the
east lobe it is at position $\approx$ 3$\arcsec$; this is in
accord with what is seen in the \mh image (Section 3.2).

\begin{figure}
\epsscale{0.45}  \plotone{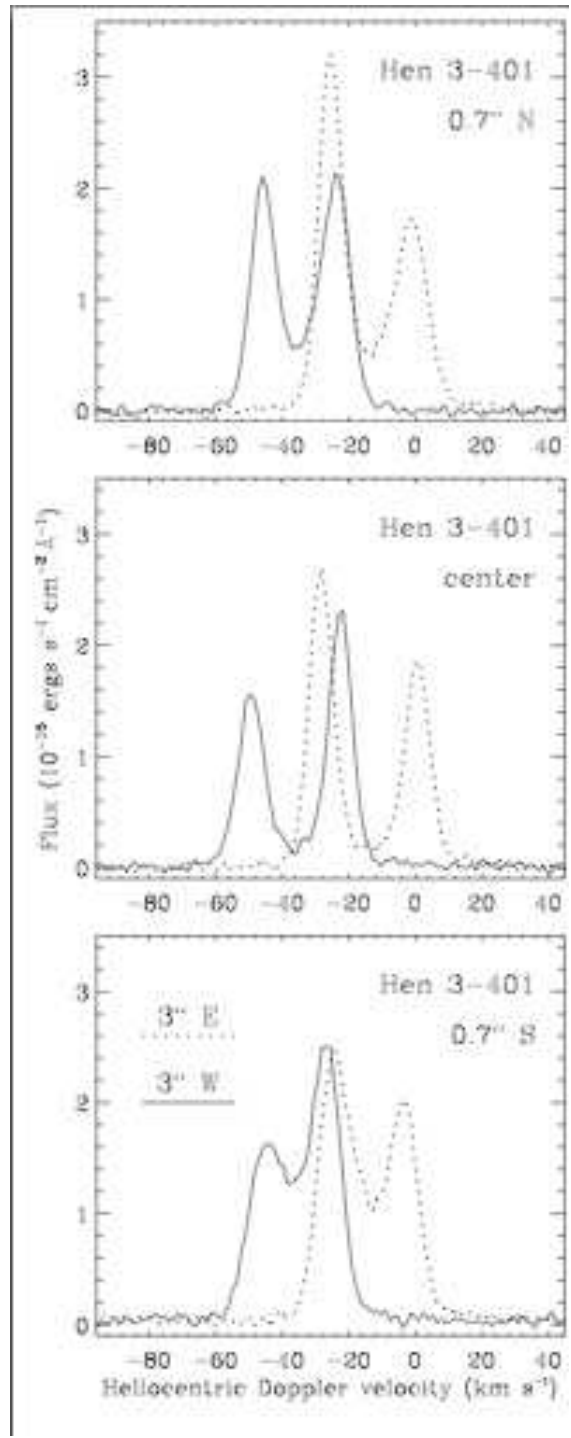}
%\plotone{traceHen.eps} 
\caption{Slices in velocity at slit positions $+3\arcsec$ and $-3\arcsec$ from the central star and
width 0.26'' across the \mh position-velocity diagram of Hen
3-401; see Fig. \ref{spec_hen3} for reference. \label{slice_hen3} }
\end{figure}

Slices in velocity across the positions $+3\arcsec$ and
$-3\arcsec$ with width 0$\farcs$26 are displayed in Figure
\ref{slice_hen3}. Velocity widths of the arcs are narrower along
the axis (Figure \ref{spec_hen3}b) than in the offset positions.  
This is due to projection
effects in a Hubble flow. For a hollow cylinder with walls of
constant thickness, the path length through the walls reaches a
minimum for a sight line passing through the polar axis,
translating to the smallest spread in velocity.

\placefigure{slice_hen3}

These observations reveal a radial velocity for the system of
V$_{\rm hel}$(\mhns) = $-$24 km~s$^{-1}$ (V$_{\rm LSR}$=$-$36 km
s$^{-1}$), with V$_{\rm exp}$ = 15 km~s$^{-1}$.  These are in good
agreement with the millimeter-wave CO $\it J$=1$-$0 emission-line
values of V$_{\rm LSR}$(CO) = $-$36 km~s$^{-1}$ with V$_{\rm exp}$
= 16 km~s$^{-1}$ by \citet{loup90} (although \citet{buj91}
measured the somewhat different value of V$_{\rm LSR}$(CO) = $-$28
km~s$^{-1}$ with V$_{\rm exp}$ = 15 km~s$^{-1}$ using the CO $\it
J$=1$-$0 and CO $\it J$=2$-$1 lines).

\subsubsection{Rob 22}

\begin{figure}
\epsscale{1.0}  \plotone{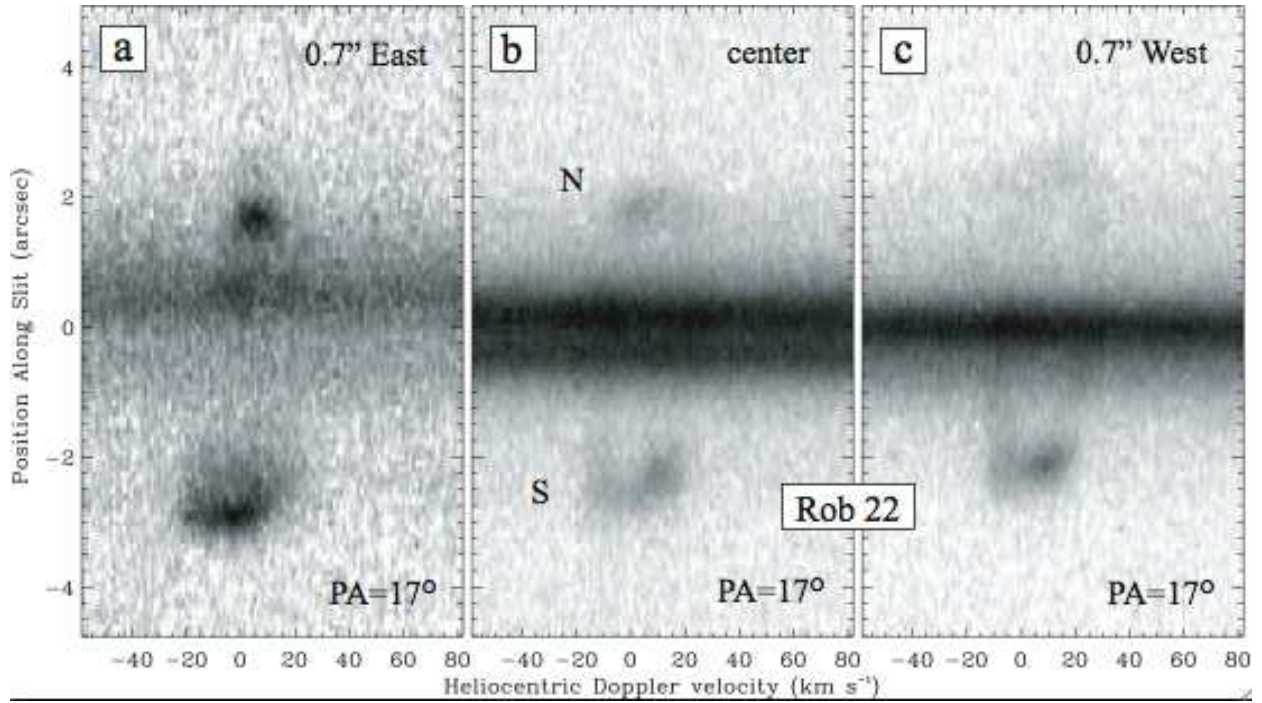}
%\plotone{fig.2Drob.eps} 
\caption{The \mh position-velocity diagram
of Rob 22.  Panels {\it a} - {\it c} correspond to the three slit
positions shown in Fig. \ref{h2_images}.  Note that the gray-scale
in each panel has been adjusted arbitrarily based upon the
strongest emission levels; since the eastern offset position ({\it
a}) does not contain a strong continuum source, the lower level
emission in the lobes is enhanced in that panel relative to the
others. \label{spec_rob22} }
\end{figure}

Figure \ref{spec_rob22} shows the long-slit Phoenix spectra of
Rob~22, with one slit positioned along the middle of the lobes
through the center of the nebula (panel {\it b}) and offset slit
positions on either side. The very weak \mh emission and the
complex continuum structure in the long-slit spectra indicate that
even a narrowband F212N image is actually dominated by dust
structures seen in reflected continuum light. This is true over
most of the nebula, with the exceptions being the N and S ends of
the polar lobes.  In the center slit position ({\it b}) there
appear to be two continuum sources separated by $\sim$0$\farcs$6,
with the northern one the brighter, while in the slit position to
the west there is one continuum source. These are consistent with
the \mh image. 
These continuum sources likely represent light from the obscured 
central star scattered toward us by regions of enhanced density.
The three panels show a slight positional offset in
the emission pattern, in agreement with the ``S'' shape of the
nebula.

\placefigure{spec_rob22}

The spectra of the nebula show very faint \mh emission from two
hollow expanding lobes with an ellipsoidal velocity structure.  
The southern lobe is brighter and larger
in \mhns, in agreement with the \mh image.  The nebula appears to
have a polar axis very close to the plane of the sky, although the
southern lobe is slightly blue-shifted and thus tilted slightly
toward us.  The lateral expansion speed of the lobes (difference
in velocity between the front and back of the S lobe) is
approximately 30 km s$^{-1}$.  Thus V$_{\rm exp}$ $\approx$ 15 km
s$^{-1}$ and V$_{\rm hel}$(\mhns) $\approx$ +5 km~s$^{-1}$, which
transforms to V$_{\rm LSR}$ $\approx$ $-$7 km s$^{-1}$. These are
in general agreement with the OH maser measurement of V$_{\rm
LSR}$(OH) = $-$6$\pm$3 km s$^{-1}$, V$_{\rm exp}$=20 km s$^{-1}$
\citep{allen80,silva93}.  However, they differ from the
molecular-line CO $\it J$=1$-$0 measurement of V$_{\rm
LSR}$(CO)=$-$0.1 km s$^{-1}$, V$_{\rm exp}$ $\sim$ 35 km s$^{-1}$
\citep{buj91}, although the investigators commented on the severe
galactic CO contamination in this direction.

\section{DISCUSSION}
\label{discussion}

\subsection{IRAS 16594$-$4656}

\begin{figure}
\epsscale{1.0}  \plotone{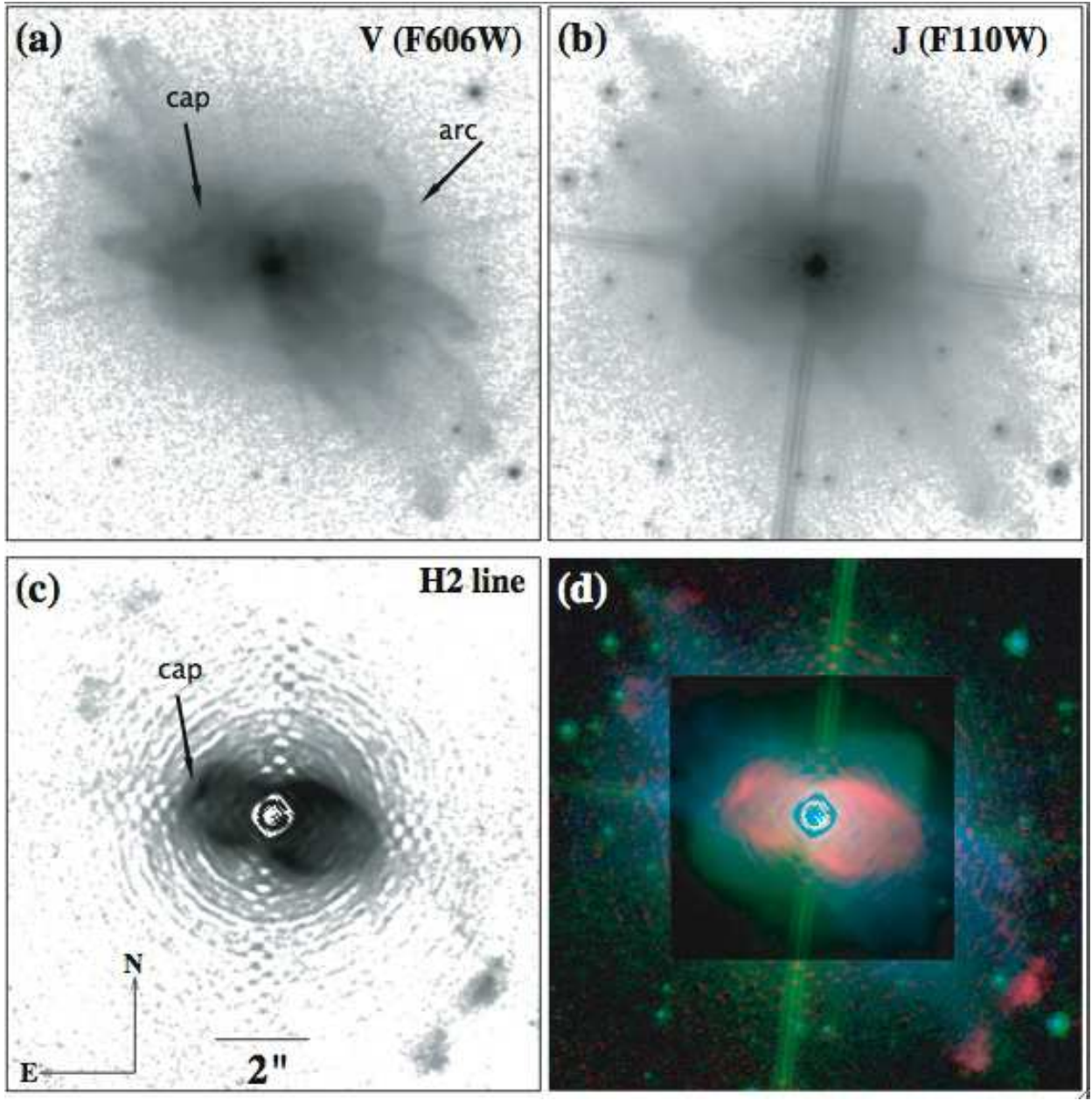}
%\plotone{i16594-f606_f110w_h2.eps} 
\caption{Images of IRAS
16594$-$4656 at various wavelengths: (a) F606W ($\sim${\it V}
band; $\lambda_{eff}$ = 0.603 \um, $<\Delta\lambda>$ = 0.150 \um),
(b) F110W ($\sim${\it H} band), and (c) \mh emission (from Fig.
3). Figure (d) is a color composite of the three, with blue
representing the {\it V} band, green the {\it J} band, and red the
\mh emission. To better show the details in the central region, it
is shown as an inset with the intensity scaled down by a factor of
ten and then displayed with a logarithmic scale. \label{16594_multi} }
\end{figure}

The PV diagrams for \lily indicate an expanding ellipsoidal nebula
with its major axis nearly in the plane of the sky.  This is a
different orientation and morphology than had been concluded based
upon the visual images of the nebula.  In the visual images (see
Fig. \ref{16594_multi}a), the star appears bright and the SW
lobe(s) appears brighter and perhaps larger than the NE lobe(s).  This is in
contrast to nebulae like AFGL 2688 (Egg nebula), IRAS
17150$-$3224, IRAS 17441$-$2411, and Rob 22, in which the star is
obscured and the two lobes look more symmetrical in size, suggesting 
that the polar axis is very nearly in the plane of the sky. Inspection of
the visual images had previously led us to conclude that \lily was
inclined at an intermediate orientation ($\sim$30$\arcdeg$) with respect to the plane
of the sky with the SW lobe oriented in the direction of the
observer \citep{su01}.  Even the new \mh image, with the SW
lobe appearing larger than the NE lobe, suggests this, assuming
symmetry in the lobes.

\placefigure{16594_multi}

However, other more recent observations at other wavelengths support our conclusion
based upon kinematics that the axis of the main bipolar lobes, as
delineated in the H$_2$ image, is nearly in the plane of the sky. A
polarization study of this source \citep{ueta05} indicates a
hollow shell with reflection from the walls.  The polarization
structure is aligned with the major axis of the nebula, with low
polarization perpendicular to this, suggesting the presence of a
dusty torus. A high-resolution mid-infrared study clearly shows
the limb-brightened ends of such a torus \citep{volk06}.  The
mid-infrared image does not show the elliptical structure expected
of a torus seen at an intermediate angle, but rather shows a bar
with bright ends, consistent with a torus seen nearly edge-on. The
bright mid-IR emission is located at the two ends of the pinched
waist seen in the H$_2$ image, showing the presence of enhanced
dust density which shapes the outflow. The approximately similar
amounts of polarizations seen at opposite sides of the nebula are
consistent with a nebula in which the bipolar axis is close to the
plane of the sky rather than at an intermediate angle, and the
authors suggest an angle of 15$\arcdeg$ with respect to the plane
of the sky \citep{ueta05}. Thus both the polarization and
mid-infrared images are consistent with the result of this
kinematic study, and the corresponding one of \citet{vds08}, 
that the bipolar lobes are aligned nearly in the plane of the sky.

The \mh image shows limb-brightened lobes, consistent with the
idea that they were carved out by collimated outflows.  Based on
the angular diameter across the lobes of 2$\farcs$0 and the measured V$_{exp}$= 9
km~s$^{-1}$, one can calculate an approximate expansion age for
the lobes of 530 $\times D_{kpc}$ yr.  Using the polar velocity 
determined using {\it SHAPE}, along with the angular diameter between 
the two ends of the bipolar lobes of 4$\farcs$0, leads to a expansion 
age of 740 $\times D_{kpc}$ yr
This leads to an estimated age for the lobes of 1400-1900 yr, 
using a distance of 2.6 kpc based on an assumed total luminosity 
of 10$^4$ L$_{\sun}$ \citep{hri00}.

Comparing the morphology of the nebula in different wavelength
regions shows several striking features.  In visible light (Fig.
\ref{16594_multi}a), the nebula shows a basic bipolar structure,
but with a series of three pairs of point-symmetric %filaments.  
thin lobes.  Thus at this wavelength it has been 
classified as point-symmetric and multipolar.  Portions of several
circumstellar rings are also seen \citep{hri01}.  In the J band
(Fig. \ref{16594_multi}b) the %filaments 
thin lobes
are present %but not as prominent, 
and a portion of at least one ring is seen faintly.  
%Instead one sees more prominently 
Seen more prominently, however, is a somewhat
elliptical morphology extending from the northwest to the
southeast
that is also present in the visible image.  
However, it is not actually an ellipse, since the
axes are not separated by 180$\arcdeg$; the northwest lobe is at
a position angle of 305$\arcdeg$ and the southeast one at
110$\arcdeg$. 
The published H-band image
is similar \citep{su03}, with the star relatively brighter
compared to the fainter %filaments.  
thin lobes.
The H$_2$ image (Fig.
\ref{16594_multi}c), in contrast, shows a clear bipolar,
``peanut'' morphology with shorter, more spherical lobes.
The major axis is at a position angle
of 75$\arcdeg$ with the west lobe larger.  Some of
the structures that define these peanut-shaped lobes can be seen upon close
examination in the V and H images.  Among these is the bright
``cap'' at the end of the eastern lobe. 
As mentioned earlier, the mid-IR images at 10 and 20
$\mu$m reveal two bright regions just outside the pinched waist of
the ``peanut.'' The mid-IR emission also faintly delineates the
edges of the two peanut-shaped lobes seen in H$_2$. \citet{volk06}, in comparing
the mid-IR images to this H$_2$ image, concluded that the
coincidence of the faint mid-IR emission with the H$_2$ emission
from the walls of these lobes suggested that the mid-IR emission was
likely partially shock heated.  The walls of these peanut-shaped lobes are clearly
outlined in polarized light \citep{ueta07}, agreeing with the
position of the dust seen in the mid-IR.

In a composite image (Fig. \ref{16594_multi}d), we see the
enhanced H$_2$ emission from the walls of the peanut-shaped bipolar lobes and
the H$_2$ at the ends of the %filaments.  
long, thin lobes.
The non-uniformity in the
emission from the walls may indicate lower density regions,
through which jets may have carved holes from which the %filaments
long, thin lobes
may have emerged.  This composite image also shows clearly
several different structures seen in this object at different
wavelengths, reinforcing the idea that the morphology of this
nebula is more complex than simply symmetric bipolar lobes.  It is
more complex than either of the other PPNs of this study.

In their excellent study  of this object, \citet{vds08} used their long-slit H$_2$
spectra together with our spectra (from the archive) to produce channel maps and from these to 
determine a more detailed three-dimensional image of the shell structure 
of IRAS 16594$-$4656.  While it basically consists of two hollow bipolar 
lobes joined at the waist, it is more complex, with some holes and extended structures
(see their Fig. 8).  These correspond to the brightness variations and the ``cap'' seen 
in our H$_2$ image.  They also analyzed [OI] and [CI] emission lines, 
and making the reasonable assumption that these lines were produced in the same shock 
as the H$_2$ lines, they used these as diagnostics of the physical conditions in the 
shock region.  From these, they deduce that the emission arises in very high 
density regions, 3 $\times$ 10$^6$ $\le$ n$_e$ $\le$ 5 $\times$ 10$^7 c$m$^{-3}$.
At these conditions of density, temperature \citep[T$_{rot}$$\approx$1400 K;][]{vds03}, 
and velocity (8 km s$^{-1}$), the excitation can be due to either C-type or J-type shocks \citep{vds08}.  
They also made long-slit observations of the [FeII] 1.645 $\mu$m emission line, 
which indicate that this emission arises, not in the lobes, but from a wind close to the star.

%In their lower resolution spectroscopic study of this object,
%\citet{vds03} observed a dozen \mh lines and determined that that
%\mh was excited by C-shocks.  In a more recent study,
%\citet{vds08} also obtained long-slit spectra similar to ours for
%this object, although at different angles.  Their H$_2$ PV
%diagrams look similar to ours.  They also obtained long-slit
%spectra in the lines of [FeII] 1.645 $\mu$m and H Pa$\beta$, from
%which they found that the [FeII] lines originate close to the
%star, in contrast to the H$_2$.  The [FeII] lines are attributed
%to a wind close to the central star.  Such a wind is also
%supported by the P~Cyg-profile seen in their H$\alpha$ spectrum,
%which indicate a velocity of $\sim$80 km~s$^{-1}$ with wings up to
%1200 km~s$^{-1}$.  They use the spectral-kinematic information to
%construct a 3D image of the H$_2$ shell, which documents the
%complexity of the morphology.  Their recent study and our present one 
%are mutually complementary in analyzing this object.

\subsection{Hen 3-401}

This nebula is very highly collimated, with the lobes extending
$\sim$14$\farcs$5 from the central star with a length to width
ratio for each lobe of $\sim$7. The lobes in this object do not appear to be
closed, but rather have ``tattered'' ends.  The PV diagrams show a
Hubble flow in the velocity in the lobes. This can arise due to an
eruptive ejection (ballistic motions) or (self-similar)
hydrodynamic growth patterns \citep{balick02}. The velocities show
that the eastern lobe is tilted away from us and the western lobe
toward us.  From the complementary image and spectroscopy, an
estimate can be made of the velocity at the ends of the lobes and
the inclination angle of the bipolar axis.  With the measured
length of the open-ended lobes of 14$\farcs$5 and the width of
2$\farcs$2 \citep[from the sharpened V image,][]{sahai99b}, and the
lateral lobe expansion velocity of 15.5 km s$^{-1}$, one can
calculate a minimum expansion velocity at the ends of the lobes of
$\sim$200 km s$^{-1}$. The PV diagram along the bipolar axis (Fig.
\ref{spec_hen3}b) shows that the velocity at the ends of the two lobes differ by 20
km s$^{-1}$ from that of the star. This implies an inclination
angle of $\leq$ 6$\arcdeg$ to the plane of the sky.  
This is somewhat smaller than we found from the appearance of the PV diagram
in Section~\ref{hen3 PV}, but both clearly indicate a small inclination angle.

\begin{figure}
\epsscale{0.9}  \plotone{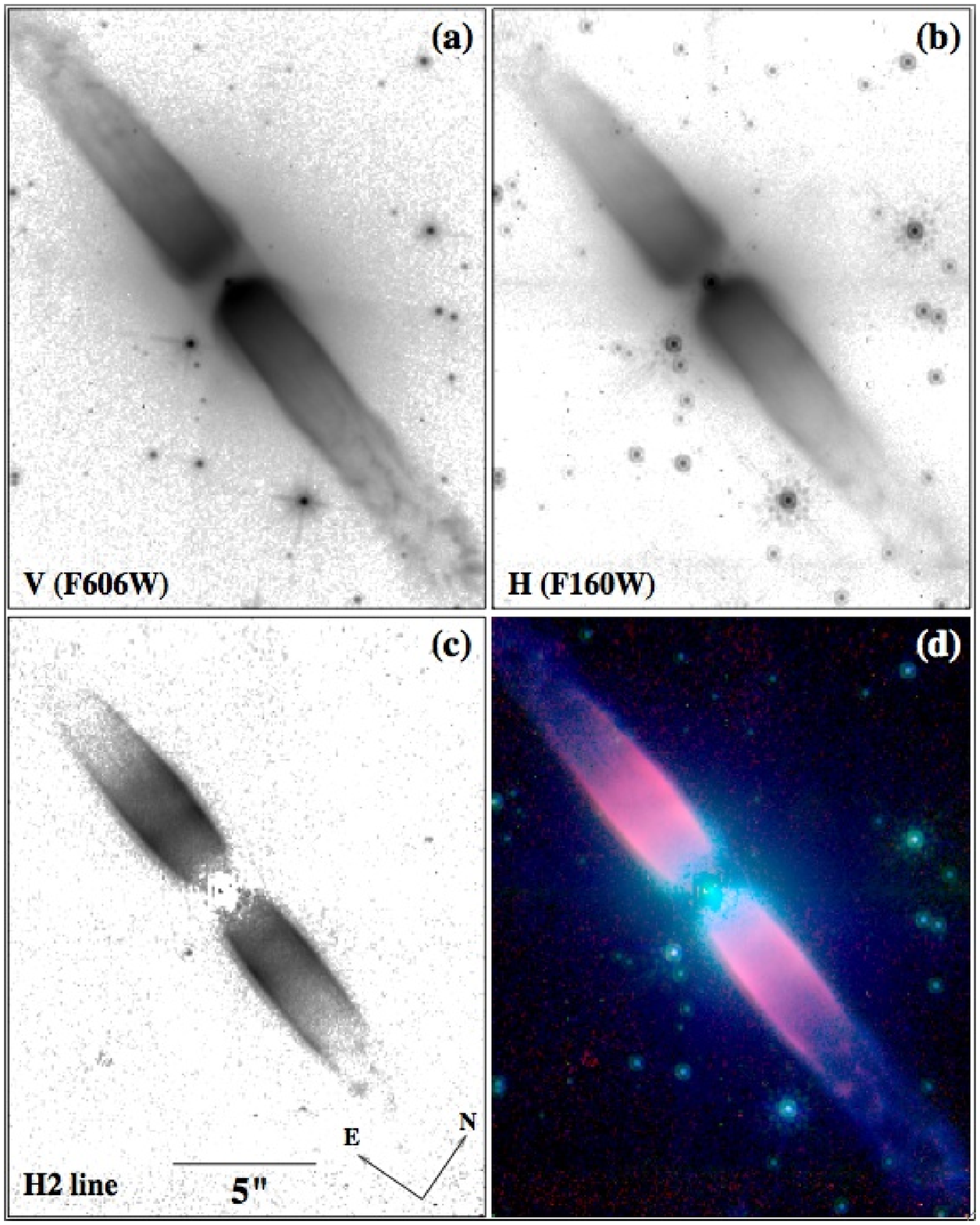}
%\plotone{hen401-f606_f160w_h2.eps} 
\caption{Images of Hen~3$-$401
at various wavelengths: (a) F606W ($\sim${\it V} band;
$\lambda_{eff}$ = 0.603 \um, $<\Delta\lambda>$ = 0.150 \um), (b)
F160W ($\sim${\it H} band), and (c) \mh emission (from Fig. 3).
Figure (d) is a color composite of the three, with blue
representing the {\it V} band, green the {\it H} band, and red the
\mh emission. \label{hen3_multi} }
\end{figure}

In visible light, a ``skirt'' structure is seen surrounding the
inner parts of the bipolar lobes and extending coaxially with the
lobes for $\sim$3$\arcsec$ from the star \citep[discussed by][]{sahai99b}.  
This skirt also appears
in the H-band image but not in H$_2$. This is seen in Figure
\ref{hen3_multi}.  The western lobe is the brighter in scattered light, with
brightness ratios in the different bands of 2.4 
\citep[V;][]{sahai99b}, 1.4 (J), and 1.4 (2.15 \umns). 
If one assumes an intrinsic symmetry in the two
lobes, then this suggests some obscuration by an extended
equatorial disk.  
In contrast, the eastern lobe is brighter in  \mhns, with a ratio of 1.2. 
This may be a result of the differing local shock conditions in the nebula.
%However, in  \mh the eastern lobe is brighter, with a ratio of 1.2.  
%We do not have a good explanation for this apart from an asymmetry in
%the nebula.
There is a suggestion of a torus from the
optical image \citep{sahai99b} and from the polarization
measurements \citep{scar95,ueta07}.   A recent high-resolution
mid-infrared imaging study shows that the mid-IR emission arises
mostly from a 1$\arcsec$ core, with weaker emission arising from
the walls extending out to about 3$\arcsec$ from the star
\citep{muthu06}. These authors suggest that the core contains a
flared disk, based upon the temperature gradient detected in the
core.

\placefigure{hen3_multi}

In the H band, the star is more prominent than in the visible and
the inner region of the lobes is slightly less prominent.  In
visible light, there is a bright inner edge to the lobes, with
enhanced light out to the ends of the skirt.  This is also seen in
the H band but not the H$_2$ image.  A faint halo is seen in both
the visible and H-band images.  The H$_2$ emission is particularly
bright 2.0$-$5.5$\arcsec$ from the star. These different regions
of brightness are seen in the composite image (Fig.
\ref{hen3_multi}d).

Based upon the extremely large intensity ratio of the \mh 1$-$0
S(1) to 2$-$1 S(1) line of 26 \citep{garher02}, it is clear that
the excitation mechanism is due to shocks.  These presumably arise
as the fast wind, that carved out the narrow lobes, produces shocks in
the walls that excite the \mhns.  H$\alpha$ wings are observed to extend out
to $\sim$1600 km s$^{-1}$ \citep{garlar99b}.

The young PN M 2-9 is very similar to Hen 3-401 in visible
appearance, as has been remarked upon by \citet{garlar99b}, and
also in its mid-infrared appearance \citep[although with not quite as 
pinched a waist;][]{smith05a} and its kinematics. 
It possesses smooth, almost cylindrical lobes,
a bright central star, and a thin \mh zone
\citep{hora94,smith05b}. A recent study of winds driven by
magnetic pressure from a toroidal field was able to reproduce well
the morphology of these two objects \citep{garseg05}.

\subsection{Rob 22}

The PV diagrams for \rob show an expanding nebula with the \mh
emission arising from the ends of the lobes and the walls.  The
bipolar axis lies nearly in the plane of the sky with the south
lobe expanding slightly toward us compared with the north lobe.
Given the expansion velocity of $\sim$15 km~s$^{-1}$ and the
angular size of the lobes of $\sim$2.0$\arcsec$, this leads to a
kinematic age of $\sim$320 $\times D_{kpc}$ yr.  For a distance of
$\sim$2 kpc \citep{allen80}, this yields an age for the lobes of
$\sim$640 yr.

\begin{figure}
 \epsscale{1.0}  \plotone{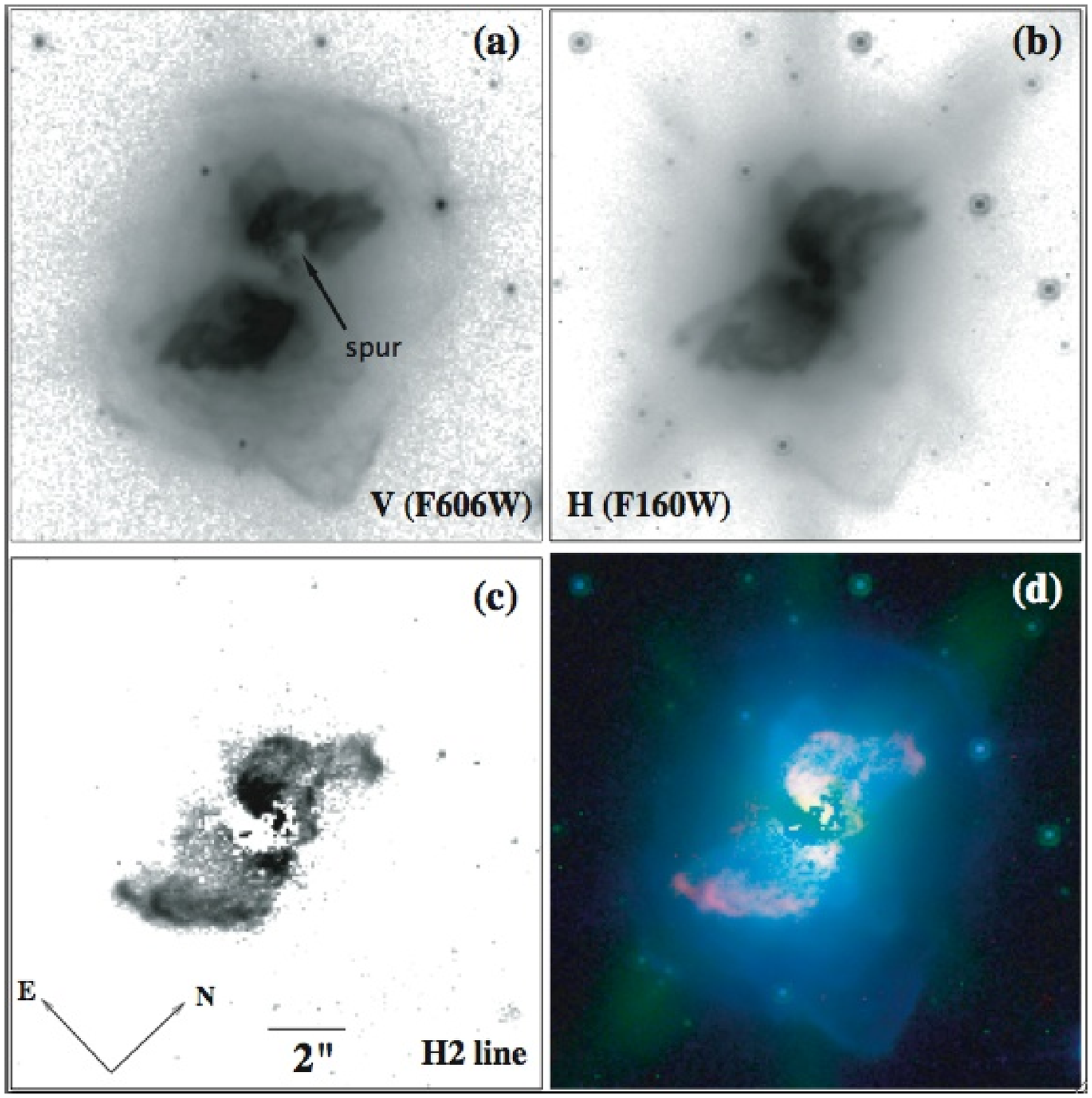}
%\plotone{rob22-f606_f160w_h2.eps}
\caption{Images of Rob~22 at
various wavelengths: (a) F606W ($\sim${\it V} band;
$\lambda_{eff}$ = 0.603 \um, $<\Delta\lambda>$ = 0.150 \um), (b)
F160W ($\sim${\it H} band), and (c) \mh emission (from Fig. 3).
Figure (d) is a color composite of the three, with blue
representing the {\it V} band, green the {\it H} band, and red the
\mh emission. \label{rob22_multi}}
\end{figure}

In the visual image \citep[Figure \ref{rob22_multi}a; ][]{sahai99a},
the south lobe is brighter than the north lobe with a ratio of 1.6.
In the near-infrared, the north lobe is slightly brighter, with ratios 
of 1.05 in H (F160W; Figure \ref{rob22_multi}b), 1.1 in 2.15 $\mu$m, and 
1.2 in the \mh image (Figure \ref{rob22_multi}c). 
If the reduced relative brightness in the south lobe in the V band is due to
extinction by the dense equatorial disk, then the disk would need
to be extremely large if it were orthogonal to the axis of the
outflow, which as we have seen is very close to the plane of the
sky.
%({\it i} $\le$ ?$\arcdeg$).
It may be more likely that there is simply an intrinsic asymmetry
in the lobes that contributes most of the difference.

\placefigure{rob22_multi}

If we compare the different images (Fig. \ref{rob22_multi}), we
see a strong point symmetry in the ``S'' shape.  Even at 7.7 and
8.7 $\mu$m this ``S'' shape is seen \citep{muthu04}.  Such a
pattern can be formed by collimation of an outflow by a precessing
disk. The H$\alpha$ emission has wings to at least $\pm$450
km~s$^{-1}$ \citep{allen80}.

The V image shows a distinct dark band separating the lobes.  In
the H image, this dark lane is not so prominent, and a bright, compact 
clump appears.  This bright clump is red but not point-like, and likely
represents warm dust heated by the obscured central star.
The nebula is brighter close to the star, with a
particularly bright region NE of the center of the nebula. A
similar morphology to the H-band image is seen in a recent
adaptive optics very high resolution (R$\sim$0.06$\arcsec$) K-band
image obtained by \citet{lag08}; this shows some finer detail and
in particular a bright linear extension toward the SE from the
very bright region near the center, defining the southern part of
the ``V''-shaped central obscured region. The halo is evident in
both the V and H images. A faint region in the N lobe (referred to
as a ``spur'' by \citet{sahai99a}) is seen in both the V and H
images, although in the V image it is outlined by emission farther
out in the lobe while in the H image it is outlined by emission
closer to the center. In the \mh image, the emission is
particularly strong in the same region NE of the center, with a
less bright region on the other side (SW) of the center.  In
addition, the \mh is particularly bright along the ends of the
lobes.  This can arise due to shock excitation produced as the collimated 
wind, that may have produced the S-shaped nebula,
interacts with the AGB wind of the halo.  Polarization
measurements show scattering from the ends of the lobes, in
addition to the presence of a equatorial dust torus
\citep{ueta07}.

The faint region ``spur'' in the N lobe is evident in all three
images.  It is interesting that in the OH observations
\citep{dyer01}, the more negative velocities arise from this faint
region; perhaps it represents an evacuated hole punched through
the nebula.

The OH emission appears to be confined primarily to the equatorial
regions \citep{sahai99a,dyer01,zij01}, where it may represent the expansion
or rotation of a disk.  The mixed C and O circumstellar chemistry
might also suggest a disk, as this explanation has been advocated
as the cause of the mixed chemistry in several other post-AGB
objects, such as the Red Rectangle \citep{win03}.  The detection
of hot dust, manifested in the spectral energy distribution of this object 
(but not in the other two), is consistent with a disk. Or it may be that this 
object is in transition between an O-rich chemistry to a C-rich one, with 
the PAH emission arising closer to the star, as may be the case for PN
with [WC] central stars.

\subsection{Comparison With Other Spatial-Kinematic Studies of PPNs}

A small number of other bipolar PPNs have also been studied
spatially or kinematically in \mh.  We discuss these for
comparison.

AFGL 2688 (``Egg'' nebula) has been imaged using {\it HST}-NICMOS
\citep{sahai98} and shows \mh emission from clumps in its
spindle-shaped closed lobes, especially near their outer ends.
This is consistent with the emission arising along the walls of
the lobes. It also shows strong emission from an extended
equatorial region. A kinematic study by \citet{kastner01} shows
that the highest velocity \mh along the polar axis resides close
to the star and that the velocity of the \mh {\it decreases} along
the polar axis away from the star.  They find a similar effect for
AFGL 618. They interpret this as resulting from a faster
collimated wind colliding with slower-moving AGB material. The
equatorial \mh shows one side to be blue- and the other
red-shifted, which \citet{kastner01} interpret indicating a
combination of expansion and rotation.   An alternate
interpretation would be multiple bipolar jets along the equator
\citep{cox03} rather than rotation, and this seems to be the more
likely explanation.
This decrease with distance along the polar axis seen in AFGL 2688 and
AFGL 618 is in contrast with 
what is seen in Hen 3-401.  Hen 3-401 displays a clear trend of the 
\mh velocity {\it increasing} in velocity with distance away from the star, 
in accord with what one would expect from an unhindered ejection of
material from the region of the central star.

\citet{davis05} recently made spatially-resolved high-resolution
(R$\sim$20,000 or 16 km~s$^{-1}$) \mh observations of four bipolar
PPNs.  They observed them with the spectrograph slit oriented
along the bipolar axis and then offset slightly to each side,
similar to our observing pattern for Hen~3$-$401 and Rob~22.  In
two of these objects, M~1$-$2 and IRAS 17441$-$2411, one sees a
distinct velocity difference between the two lobes ($\Delta$V of
$\sim$55 and $\sim$10 km s$^{-1}$, respectively), for IRAS
17243$-$1755 \mh is detected only weakly from one lobe, and for
IRAS 17150$-$3224 the two lobes have similar velocities (bipolar
axis in the plane of the sky). For this last object, they also
suggested that there may be rotation in the lobes based on
systematic velocity differences seen in the offset positions.

\cite{hri06} also studied IRAS 17150$-$3224 with higher spectral
resolution, and they obtained a high-resolution (0$\farcs$2) {\it
HST}-NICMOS \mh image.  The image shows the \mh emission to come
primarily from clumps near the ends of the lobes, but these are
not resolved in the spectra, which basically show the integrated
emission and velocity of the clumps.  The two lobes have a similar
velocity, as in the above study, with slight velocity differences
in the offset positions due to small effects of the different \mh
clumps rather than rotation.

As we have referred to previously, \citet{vds08} also recently studied 
the kinematics of IRAS 16594$-$4656.  They also observed with
the Phoenix spectrograph on Gemini-South at even somewhat higher 
spectral and spatial resolution than we did.  Their spectra show a 
similar picture of an expanding H$_2$ ellipsoid with V$_{exp}$ = 8 km
s$^{-1}$.  They combined their spectra with ours to produce a 
detailed three-dimensional image of the shell structure.
 
In none of these earlier studies %, except for that of \citet{vds08}, 
does one see such clearly-resolved
spatial-kinematic structures as are seen in the new Gemini-South
Phoenix data of IRAS 16594$-$4656, Hen~3$-$401, and Rob~22.  In
all three the emission from the walls of the lobes is clearly
traced in the PV diagrams.  The small slit size and high spectral
resolution, together with the high spatial resolution images, 
result in exceptional kinematic detail for a PPN.

\section{SUMMARY AND CONCLUSIONS}

The high-resolution {\it HST}-NICMOS \mh images of these three
PPNs have allowed us to determine the spatial location of their
\mh emission and the spatially-resolved \mh spectroscopic
observations have allowed us to determine the kinematics of the
\mh emitting regions.  In all three cases, the systemic and
expansion velocities are similar to those determined from the
molecular-line CO or OH measurements.  We find the following
results for these three PPNs.

{\it IRAS 16594$-$4656}: While the V and H-band images show a
complex multi-lobe structure, the \mh images shows a clear
bipolar, peanut shape, although with variations in density (``holes'') 
and structure \citep[see][]{vds08}.  The \mh emission originates along the
walls of the lobes (sides and ends), with fainter emission from
more distant (ejected?) clumps.  The PV diagram shows the \mh to
arise in an expanding ellipsoidal velocity structure, which is in
contrast to the bilobes of the density structure.  The kinematics
indicates that the bipolar lobes are nearly in the plane of the
sky (i$\approx$10$\arcdeg$); this differs from the earlier interpretations of the lobes as
being at some intermediate orientation, but it is consistent with
recent mid-IR imaging and near-IR polarization studies.  The lobes
are estimated to have an age of $\sim$1600 yr.

{\it Hen 3-401}: The \mh emission originates from the sides of
lobes, which have open ends.  The lobes are tilted somewhat to the
plane of the sky ($\sim$10$-$15$\arcdeg$), with the western lobe
moving toward us, and they show an increasing velocity with radial
distance (Hubble flow). An estimated age for the lobes is
$\sim$1100 yr.
The open ends and unhindered outflow may be the consequence of 
its higher degree of collimation (greater linear momentum). 
%and the more evolved state of the central star and thus the system.  No - since the age for the nebula is less than for 16594.

{\it Rob 22}: The \mh emission originate primarily from ends of
the S shaped nebula and from regions of the S shape near the
obscured central star.  The nebula is nearly in the plane of the
sky, consistent with the absence of a visible star due to
obscuration by a disk.  An age of $\sim$630 yr is estimated for
the lobes.

\mh surveys of PPNs have shown that \mh emission is commonly found
in those with a bipolar morphology.  As can be seen from this
study, the combination of \mh high-resolution images and
spatially-resolved, high-resolution spectra provides valuable
insight into the structure and shaping mechanisms for these
bipolar nebulae.

\acknowledgments

We thank Steve Ridgway, Bernadette Rogers, Kevin Volk, and Claudia
Winge for making the Phoenix queue observations, Ken Hinkle for
assistance in planning the Phoenix observing programs, and Anibal
Garc\'{i}a-Hern\'{a}ndez and Griet Van de Steene for making
available to us their medium-resolution 2 $\mu$m spectra in
digital form. We thank Nico Koning for making the SHAPE images 
and Wenxian Lu for help with the image measurements.
The comments of the referee were helpful in improving the presentation.
We acknowledge grants from NASA that provided partial support for 
B.J.H. (GO-07840.02-A, GO-09366.01-A), 
R.S. (GO-07840.01-A ,GO-09463.01-A, GO-09801.01-A), 
N.S. (HF-01166.01A), and K.Y.L.S. (GO-09366.03-A)
from the Space Telescope Science Institute, which is
operated by the Association of Universities for Research in
Astronomy, Inc., under NASA contract NAS5-26555. B.J.H. also
acknowledges the support of the National Science Foundation under
Grant No. 0407087 and R.S. thanks NASA for partially funding this 
work by a NASA LTSA award (no. 399-20-40-06).
Some of the research described in this paper was carried out by R.S. at the
Jet Propulsion Laboratory, California Institute of Technology, under a 
contract with the National Aeronautics and Space Administration.
This research made use of the SIMBAD database, operated at CDS,
Strasbourg, France, and NASA's Astrophysics Data System.

\clearpage

%\tablenum{1}
\begin{deluxetable}{lclccccc}
\tablecaption{{\it HST} NICMOS Observing Log\label{obslog}}
\tablewidth{0pt} \tablehead{ \colhead{IRAS ID}&\colhead{Other
Names}&\colhead{Observation}& \multicolumn{5}{c}{Total Exposure
Times (s)\tablenotemark{a}}\\ \cline{4-8}
\colhead{}&\colhead{}&{Date}&\colhead{F212N}&\colhead{F215N}&\colhead{F110W}&\colhead{F160W}&\colhead{F222M}}
\startdata
10178$-$5958 & Hen 3-401       &1998 Mar  6&1728 &1728 &\nodata &576 &1152 \\
10197$-$5750 & Rob 22        &1998 May 18&1344 &1344 &\nodata &320 &768 \\
16594$-$4656 & \nodata      &2002 Aug 13&1215 &1215 &192 &\nodata &\nodata\\
\enddata
\tablenotetext{a}{Note that in two of our previous papers involving NICMOS images \citep{su03,hri06}, we incorrectly listed as the total exposure times values that were one-third of the correct values. Thus for our F212N and F215N images of IRAS 17150$-$3224, for example, the total exposure times were 2448 s, not 816 s as listed \citep{hri06}. }
\end{deluxetable}

\clearpage

%\tablenum{2}
\begin{deluxetable}{lcccccc}
\tablecaption{Photometric Calibration Values\tablenotemark{a}
\label{obscal}} \tablewidth{0pt} \tablehead{
\colhead{Filter}&\colhead{$\lambda_{c}$}&\colhead{$\Delta
\lambda$}
&\colhead{PHOTFLAM}&\colhead{PHOTFNU}&\colhead{ZP(Vega)}&\colhead{PSF
  FWHM\tablenotemark{b}} \\
\colhead{      }&\colhead{\um}&\colhead{\um}& \colhead{erg
 cm$^{-2}$\umns$^{-1}$DN$^{-1}$}&\colhead{Jy sec DN$^{-1}$}& \colhead{Jy}&\colhead{\arcsec} \\
} \startdata
F212N-new&  2.1213 & 0.0087 & 1.9135E-14 & 2.8722E-5 & 664.7 & 0.19 \\
F215N-new&  2.1487 & 0.0079 & 2.1031E-14 & 3.2389E-5 & 645.1 & 0.20 \\
F212N-old&  2.1213 & 0.0088 & 2.4272E-14 & 3.6432E-5 & 664.7 & 0.20 \\
F215N-old&  2.1487 & 0.0079 & 2.5533E-14 & 3.9323E-5 & 645.1 & 0.20 \\
F110W-new&  1.1235 & 0.1630 & 2.9059E-15 & 1.2234E-6 &1784.4 & 0.12 \\
F160W-old&  1.6060 & 0.1177 & 2.3956E-15 & 2.0610E-6 &1040.7 & 0.14 \\
\enddata
\tablenotetext{a}{The photometric calibration values were taken
from the online NICMOS photometry performance for post-NICMOS
(new) cooling System installation (operating temperature of
detector $\sim$77K) and pre-NICMOS (old) Cooling System
installation (operating temperature of detector $\sim$62K). }
\tablenotetext{b}{Based upon a Gaussian fit to the observed field
stars.  }
\end{deluxetable}

\clearpage

%\tablenum{3}
\begin{deluxetable}{lcrccc}
\tablecaption{High-Resolution \mh Spectroscopic Observing Log
\label{ph_obs}} \tablehead{ \colhead{Object} &\colhead{Date}
&\colhead{P.A. ($\arcdeg$)\tablenotemark{a}}&\colhead{Position}
&\colhead{t$_{exp}$ (s)\tablenotemark{b}}&\colhead{No. Obs.} }
\startdata
Rob 22 & 2003 Jan 14 &   17 & through star          & 500 & 2 \\
Rob 22 & 2003 Jan 14 &   17 & 0$\farcs$7 W of star  & 500 & 2 \\
Rob 22 & 2003 Jan 14 &   17 & 0$\farcs$7 E of star  & 500 & 2 \\
Hen 3-401 & 2003 Dec 17& 73 & through star & 500 & 2 \\
Hen 3-401 & 2003 Dec 17& 73 & 0$\farcs$7 S of star & 500 & 2 \\
Hen 3-401 & 2003 Dec 17& 73 & 0$\farcs$7 N of star & 500 & 2 \\
Hen 3-401 & 2003 Dec 17& 163 & through star & 500 & 2 \\
IRAS 16594$-$4656 & 2003 May 08&  52 (b) & through star & 600 & 2 \\
IRAS 16594$-$4656 & 2003 May 08&  33 (c) & through star & 600 & 2 \\
IRAS 16594$-$4656 & 2003 May 08& 345 (e) & 0$\farcs$9 W of star & 600 & 2 \\
IRAS 16594$-$4656 & 2003 May 15& 345 (d) & through star & 600 & 3 \\
IRAS 16594$-$4656 & 2003 May 15&  72 (a) & through star & 600 & 2 \\
\enddata
\tablenotetext{a}{The letters in parenthesis for IRAS 16594$-$4656
refer to the slit orientations as displayed in Figure
\ref{spec_16594}. } \tablenotetext{b}{Exposure times per
observation. }
\end{deluxetable}

\clearpage

\clearpage

\clearpage

\clearpage

\clearpage

\clearpage

\clearpage

\clearpage

\clearpage

\clearpage

\clearpage

\clearpage

\clearpage

\begin{thebibliography}{}

\bibitem[Allen(1978)]{allen78}
    Allen, D.A. 1978, \mnras, 184, 601
\bibitem[Allen, Hyland, \& Caswell(1980)]{allen80}
    Allen, D.A., Hyland, A.R., \& Caswell, J.L. 1980, \mnras, 192,
    505
\bibitem[Balick(1987)]{balick87}
    Balick, B. 1987, \aj, 94, 671
\bibitem[Balick \& Frank(2002)]{balick02}
    Balick, B., \& Frank, A. 2002, \araa, 40, 439
\bibitem[Bragg et al.(1982)]{bra82}Bragg, S.L., Brault, J.W., \& Smith, W.H.\ 1982, ApJ, 263, 999
\bibitem[Bujarrabal \& Bachiller(1991)]{buj91}
    Bujarrabal, V., \& Bachiller, R. 1991, \aap, 242, 247
\bibitem[Cox et al.(2003)]{cox03} Cox, P., Huggins, P. J., Maillard, J.-P.,
    Muthu, C., Bachiller, R., \& Forveille, T. 2003, \apj, 586, L87
\bibitem[Davis et al.(2005)]{davis05}
    Davis, C.J., Smith, D.M., Gledhill, T.M., \& Varricatt, W.P. 2005, \mnras,
    360, 104
\bibitem[Dyer et al.(2001)]{dyer01}
    Dyer, K.K., Goss, W.M., \& Kemball, A.J. 2001, \aj, 121, 2743
\bibitem[Garc\'{i}a-Hern\'{a}ndez et al.(2002)]{garher02}
    Garc\'{i}a-Hern\'{a}ndez, D.A., Manchado, A., Garc\'{i}a-Lario, P.,
    Dom\'{i}ngueq-Tagle, C., Conway, G.M., \& Prada, F. 2002, \aap, 387,
    955
\bibitem[Garc\'{i}a-Lario et al.(1999a)]{garlar99a}
    Garc\'{i}a-Lario, P., Manchado, A., Ulla, A., \& Manteiga, M. 1999a, \aap, 513,
    941
\bibitem[Garc\'{i}a-Lario et al.(1999b)]{garlar99b}
    Garc\'{i}a-Lario, P., Riera, A., \& Manchado, A. 1999b, \aap, 526, 854
\bibitem[Garc\'{i}a-Segura et al.(2005)]{garseg05}
    Garc\'{i}a-Segura, G., L\'{o}pez, J.A., \& Franco, J. 2005, \apj,
    618, 919
\bibitem[Hinkle et al.(1998)]{hinkle98}
    Hinkle, K. H., Cuberly, R. W., Gaughan, N. A., et al. 1998,
    Proc. SPIE, 3354, ed. A.M. Fowler, 810
\bibitem[Hora \& Latter(1994)]{hora94} Hora, J. L., \& Latter, W. B. 1994, \apj, 437, 281
\bibitem[Hrivnak et al.(2004)]{hri04}
    Hrivnak, B. H., Kelly, D. M., \& Su, K. Y. L. 2004, in ASP Conf. Ser. 313, Asymmetric Planetary Nebulae
    III, ed. M. Meixner, J.H. Kastner, B. Balick, N. Soker (San Franciso: ASP),
    175
\bibitem[Hrivnak et al.(2006)]{hri06}
    Hrivnak, B. H., Kelly, D. M., Su, K. Y. L., Kwok, S., \& Sahai, R. 2006, \apj,
    650, 237
\bibitem[Hrivnak, Kwok, \& Su(1999)]{hri99}
    Hrivnak, B. H., Kwok, S. \& Su, K. Y. L. 1999, \apj, 524, 849
\bibitem[Hrivnak, Kwok, \& Su(2001)]{hri01}
    Hrivnak, B. H., Kwok, S. \& Su, K. Y. L. 2001, \aj, 121, 2775
\bibitem[Hrivnak, Volk, \& Kwok(2000)]{hri00}
    Hrivnak, B. H., Volk, K. \& Kwok, S. 2000, \apj, 535, 275
\bibitem[Kastner et al.(2001)]{kastner01} Kastner, J.H., Weintraub, D.A. ,
    Gatley, I., \& Henn, L. 2001, \apj, 546, 279
\bibitem[Kelly \& Hrivnak(2005)]{kelly05} Kelly, D.M., \& Hrivnak, B.J., 2005,
    \apj, 629, 1040
\bibitem[Koekemoer et al.(2002)]{koekemoer02} Koekemoer, A. M., et
  al. 2002, ``HST Dither Handbook'', Version 2.0 (Baltimore: STScI)
\bibitem[Kwok (1982)]{kwok82} Kwok, S. 1982, \apj, 258, 280
\bibitem[Lagadec et al.(2008)]{lag08}
    Lagadec, E., Chesneau, O., Zijlstra, A., \& Mekarnia, D. 2008,
    in Asymmetric Planetary Nebulae IV, ed. R.L.M. Corradi, A. Manchado, , N. Soker,
    in publication
\bibitem[Loup et al.(1990)]{loup90}
    Loup, C., Forveille, T., Nyman, L.~$\AA$., \& Omont, A. 1990, \aap, 227,
    L29
\bibitem[Muthu, Kwok, \& Volk(2004)]{muthu04}
    Muthu, C., Kwok, S., \& Volk, K. 2004, Bul. AAS, 36, 1570
\bibitem[Muthumariappan, Kwok, \& Volk(2006)]{muthu06}
    Muthumariappan, C., Kwok, S., \& Volk, K. 2006, \apj, 640, 353
\bibitem[Molster et al.(1997)]{mol97}
    Molster, F., Waters, L.B.F.M., de Jong, T., Prusti, T.,
    Zijlstra, A., \& Meixner, M. 1997, IAU Symp. 180: Planetary
    Nebulae, ed. H.J. Habing, H.J.G.L.M. Lamers (Kluwer:
    Dordrecht), 361
\bibitem[Molster et al.(2002)]{mol02}
    Molster, F., Waters, L.B.F.M., Tielens, A.G.G.M., \& Barlow,
    M.J. 2002, \aap, 382, 184
\bibitem[Parthasarathy et al.(2001)]{partha01}
    Parthasarathy, M., Garc\'{i}a-Lario, P., Gauba, G. et al. 2001, \aap, 376,
    941
\bibitem[Parthasarathy \& Pottasch(1989)]{partha89}
    Parthasarathy, M., \& Pottasch, S.R. 1989, \aap, 154, L16
\bibitem[Reyniers(2002)]{rey02}
    Reyniers, M. 2002, Ph.D. Thesis, Catholic University of Leuven
\bibitem[Sahai et al.(1999b)]{sahai99b}
    Sahai, R., Bujarrabal, V., \& Zijlstra A. 1999b, \apj, 518, L115
\bibitem[Sahai et al.(1998)]{sahai98} Sahai, R., Hines, D. C., Kastner, J. H.,
    Weintraub, D. A., Trauger, J. T., Rieke, M. J., Thompson, R. I., \&
    Schneider, G. 1998, \apjl, 492, L163
\bibitem[Sahai et al.(2007)]{sahai07}
    Sahai, R., Morris, M., S\'{a}nchez Contreras, C., \& Claussen, M. 2007, \aj,
    134, 2200
\bibitem[Sahai et al.(2000)]{sahai00} Sahai, R., Su, K.Y.L., Kwok, S., Dayal,
    A. \& Hrivnak, B.J. 2000, in ASP Conf. Ser. 199, Asymmetric Planetary Nebulae
    II - From Origins to Microstructures, ed. J.H. Kastner, N. Soker, S.A. Rappaport
    (San Franciso: ASP), 167
\bibitem[Sahai \& Trauger(1998)]{sahai98a} Sahai, R., \& Trauger, J.T. 1998, \aj, 116, 1357
\bibitem[Sahai et al.(1999a)]{sahai99a}
    Sahai, R., Zijlstra A., Bujarrabal, V., \& te Lintel Hekkert, P. 1999a, \aj, 117,
    1408
\bibitem[Scarrott \& Scarrott(1995)]{scar95}
    Scarrott, S.M., \& Scarrott, R.M.J.,1995 \mnras, 277, 277
\bibitem[Silva et al.(1993)]{silva93}
    Silva, A.M., Azc\'{a}rate, I.N., Poppel, W.G.L., \& Likkel, L. 1993, \aap, 275,
    510
\bibitem[Si\'{io}dmiak et al.(2008)]{siod08}
	Si\'{o}dmiak, N., Meixner, M., Ueta, T., Sugerman, B.E.K., Van de Steene,
	G.C., \& Szczerba, R.  2008, \apj, 677, 382
\bibitem[Smith(2006)]{smith06} Smith, N. 2006, \apj, 644, 1151
\bibitem[Smith \& Gehrz(2005)]{smith05a} Smith, N., \& Gehrz, R.D. 2005, \aj, 129, 969
\bibitem[Smith et al.(2005)]{smith05b} Smith, N., Balick, B., \& Gehrz, R.D. 2005, \aj, 130, 853
\bibitem[Steffen \& L\'{o}pez(2006)]{stef06}
	Steffen, W., \& L\'{o}pez, J.A. 2006, \rmxaa, 42, 99
\bibitem[Su et al.(2001)]{su01} Su, K. Y. L., Hrivnak, B. J., \& Kwok, S. 2001, \aj, 122,
    1525
\bibitem[Su et al.(2003)]{su03} Su, K. Y. L., Hrivnak, B. J., Kwok, S., \& Sahai, R.
    2003, \aj, 126, 848
\bibitem[Ueta et al.(2000)]{ueta00} Ueta, T., Meixner, M., \& Bobrowsky, M. 2000, \apj, 528, 861
\bibitem[Ueta et al.(2005)]{ueta05}
    Ueta, T., Murakawa, K., \& Meixner, M. 2005, \aj, 129, 1625
\bibitem[Ueta et al.(2007)]{ueta07}
    Ueta, T., Murakawa, K., \& Meixner, M. 2007, \aj, 133, 1345
\bibitem[Van de Steene \& van Hoof(2003)]{vds03}
    Van de Steene, G.~C., \& van Hoof, P.~A.~M. 2003, \aap, 406, 773
\bibitem[Van de Steene et al.(2008)]{vds08}
    Van de Steene, G.~C., Ueta, T., van Hoof, P.~A.~M., Reyniers, M., \& Ginsburg, A.G.
    2008, \aap, 480, 775
\bibitem[Van de Steene, Wood, \& van Hoof(2000)]{vds00}
    Van de Steene, G.~C., Wood, P.R., \& van Hoof, P.~A.~M. 2000,
    in ASP Conf. Ser. 199, Asymmetric Planetary Nebulae II - From Origins to
    Microstructures, ed. J.H. Kastner, N. Soker, S.A. Rappaport
    (San Franciso: ASP), 191
\bibitem[van Winckel(2003)]{win03}van Winckel, H. 2003, \araa, 41, 391
\bibitem[Volk et al.(2006)]{volk06}
    Volk, K., Hrivnak, B.J., Su, K.Y.L. \& Kwok, S. 2006, \apj,
    651, 294
\bibitem[Volk et al.(2002)]{volk02}
    Volk, K., Kwok, S., Hrivnak, B.J., \& Szczerba, R. 2002, \apj,
    567, 412
\bibitem[Zijlstra et al.(2001)]{zij01}
    Zijlstra, A.A., Chapman, J.M., te Lintel Hekkert, P., Likkel, L., Comeron, F., 
    Norris, R.P., Molster, F.J., \& Cohen, R.J. 2001, \mnras, 322, 280

\end{thebibliography}
\end{document}